\pdfoutput=1

\documentclass[aps,prd,preprint,superscriptaddress,showpacs]{revtex4}
\usepackage{graphics}
\usepackage{epstopdf}
\usepackage{graphicx}
\usepackage{dcolumn}
\usepackage{bm}
\usepackage{amsmath}
\usepackage{color}

\newcommand{\be}{\begin{equation}}
\newcommand{\ee}{\end{equation}}
\newcommand{\ba}{\begin{eqnarray}}
\newcommand{\ea}{\end{eqnarray}}
\newcommand{\bfig}{\begin{figure}[t]\begin{centering}}
\newcommand{\efig}{\end{centering}\end{figure}}

\def\dm03{\hbox{$\Delta m^2_{03}$}}

\begin{document}

\title{Mass hierarchy, 2-3 mixing and CP-phase with Huge Atmospheric Neutrino 
Detectors}

\author{E. Kh.  Akhmedov}
\thanks{Also at the National Research Centre Kurchatov 
Institute, Moscow, Russia.} 
\email{akhmedov@mpi-hd.mpg.de}
\affiliation{Max-Planck-Institut fur Kernphysik, Saupfercheckweg 1, 
D-69117 Heidelberg, Germany}

\author{Soebur Razzaque}
\thanks{Present address: Department of Physics, University of Johannesburg, C1-Lab 140, PO Box 524, Auckland Park 2006, South Africa.} 
\email{srazzaque@uj.ac.za}
\affiliation{School of Physics, Astronomy and Computational Sciences, 
George Mason University, Fairfax, VA 22030, USA}

\author{A.\ Yu.\ Smirnov}
\email{smirnov@ictp.it}
\affiliation{The Abdus Salam International Centre for 
Theoretical Physics, I-34100 Trieste, Italy}

\date{\today}

\begin{abstract}

We explore the physics potential of multi-megaton scale ice or water
Cherenkov detectors with low ($\sim 1$ GeV) threshold.  Using some
proposed characteristics of the PINGU detector setup we compute the
distributions of events versus neutrino energy $E_\nu$ and zenith
angle $\theta_z$, and study their dependence on yet unknown neutrino
parameters. The $(E_\nu - \theta_z)$ regions are identified where the
distributions have the highest sensitivity to the neutrino mass
hierarchy, to the deviation of the 2-3 mixing from the maximal one and
to the CP-phase. We evaluate significance of the measurements of the
neutrino parameters and explore dependence of this significance on the
accuracy of reconstruction of the neutrino energy and direction. The
effect of degeneracy of the parameters on the sensitivities is also
discussed. We estimate the characteristics of future detectors (energy
and angle resolution, volume, etc.) required for establishing the
neutrino mass hierarchy with high confidence level.  We find that the
hierarchy can be identified at $3\sigma$ -- $10\sigma$ level
(depending on the reconstruction accuracies) after 5 years of PINGU
operation.

\end{abstract}

\maketitle

\section{Introduction}

Atmospheric neutrino studies have enormous physics potential which has
not been fully explored yet. Atmospheric neutrinos can be used (i) to
explore different effects of neutrino propagation: oscillations in
vacuum and in matter, MSW resonance of neutrino
oscillations \cite{Wolf,MS}, as well as parametric enhancement
effects \cite{param}, {\it etc.}; (ii) to determine neutrino
oscillation parameters: mass squared differences, mixing angles, CP
violation effects, mass hierarchy; (iii) to search for new physics
beyond the standard framework with 3 light neutrinos: new neutrino
states, non-standards interactions, violation of fundamental
symmetries, {\it etc.}; (iv) to perform, in principle, the oscillation
tomography of the Earth.

Among major results obtained with the atmospheric neutrinos are the
discovery of neutrino oscillations~\cite{discovery}, first
measurements of $\theta_{23}$ and $\Delta
m^2_{32}$ \cite{Kajita:2011zz} and various bounds on new
physics~\cite{GonzalezGarcia:2004wg}.

The physics potential of atmospheric neutrino studies with existing
and forthcoming experiments has been widely discussed before (see,
e.g.,~\cite{before} and references therein).  In particular, the
potential of the IceCube's DeepCore (the currently existing detector
within the inner core of IceCube with an energy threshold $E_\nu\sim
10$ GeV \cite{deepcore}) for studying neutrino oscillations has been
explored in Refs.
\cite{Mena:2008rh,Giordano:2010pr,FernandezMartinez:2010am}.  The 
first experimental results on atmospheric neutrino oscillations in 
DeepCore have been reported in Ref.~\cite{DCresults}.
  
Although at the probability level the effects of the neutrino mass
hierarchy, deviation of the 2-3 mixing from the maximal one, and
CP-violation in specific oscillation channels can be of order 1, there
are a number of factors which substantially reduce the effects at the
level of observable events.  As a result, determination of the
neutrino parameters becomes difficult if possible at all.  The main
challenges include (1) relatively low statistics, especially at
energies above a few GeV; (2) the presence of both neutrinos and
antineutrinos in the original neutrino flux and the difficulty of
experimental separation of the signals from neutrinos and
antineutrinos, especially in large (megaton scale) detectors; (3) the
existence of both $\nu_e$ and $\nu_\mu$ in the original fluxes; (4) a
significant smearing of the signal over the energies and zenith
angles, related to large uncertainties in reconstruction of the
neutrino energy and direction.

Recently, the idea has been entertained that large statistics of
events that can be collected over a wide energy range in multi-megaton
detectors with low (a few GeV) thresholds, supplemented by relatively
mild technological improvements, can alleviate or remove the
above-mentioned shortcomings~\cite{our1,our2,our3}. High statistics
will allow one to make specific selections of events with different
geometries from certain ranges of energies and zenith angles in order
to enhance the sensitivity of the experiment to various neutrino
parameters, to resolve degeneracies between those parameters, and to
reduce the effects of uncertainties of the input parameters (such as
the neutrino fluxes and experimental resolutions).  Recent
considerations and discussions of the multi-megaton ice detector PINGU
(the proposed upgrade of the IceCube detector) \cite{pingu} show that
this hope may actually be realized.

In this paper we explore the possibilities of studying the neutrino
parameters in multi-megaton scale ice or water Cherenkov detectors
with energy thresholds as low as a few GeV. We study the energy and
zenith angle distributions of events and their dependence on the
neutrino mass hierarchy (ordering), the deviation of the 2-3 mixing
from the maximal one and on the CP-phase. We identify the geometry of
the events and the kinematic regions in the ($E_\nu - \cos \theta_z$)
plane where the dependence on a specific neutrino parameter is
maximal. We compute significance of measurements of these parameters
and explore dependence of the significance on the energy and zenith
angle resolutions of the detector, i.e.\ on the accuracy of
reconstruction of the neutrino energy and direction.

In our calculations we use some provisional characteristics of the
PINGU detector, in particular the effective volume and its energy
dependence.  At the same time, we perform our analysis with an effort
to make it as much as possible independent of specific experimental
features, which are yet to be determined.

The paper is organized as follows. In sec.\ II we summarize relevant
information about the oscillation probabilities. In Sec.\ III we
present distributions of events in the $E_\nu - \cos \theta_z$ plane
as well as significance plots for determination of various neutrino
parameters.  In sec.\ IV we perform smearing of the event densities
using neutrino energy and direction reconstruction functions. In this
section we also study dependence of the significance of the
determination of the neutrino mass hierarchy on the energy and angular
resolutions (the widths of the reconstruction functions) of the
detector. In addition, we discuss here the issue of degeneracy of
neutrino parameters. Sec.\ V contains discussion of our results and
conclusions.

\section{Oscillation probabilities}

In this section we discuss dependence of the oscillation probabilities
on the neutrino mass hierarchy and on the deviation of the 2-3 mixing
from the maximal one. We consider the evolution of three neutrino
flavors $\nu_f \equiv (\nu_e, \nu_\mu, \nu_\tau)^T$ in the propagation
basis, $\nu_{prop} = (\nu_e, \tilde{\nu}_2, \tilde{\nu}_3)^T$, defined
as $\nu_f = U_{23} I_{\delta} \nu_{prop}$. We use the standard
parameterization of the PMNS mixing matrix, $U_{PMNS} = U_{23}
I_{\delta} U_{13} U_{12}$, where $U_{ij}$ is the matrix of rotations
in the $ij-$ plane and $I_{\delta} = diag(1, 1, e^{i\delta})$ is the
matrix with CP-violating phase.

We consider the neutrino energy range $E_\nu > (2 - 3)$ GeV which
includes the 1-3 resonances and parametric enhancement ridges, and
where the sensitivity to the neutrino mass hierarchy is expected to be
maximal. Also in this range one expects small CP-violation effects, so
that their degeneracy with the effects of the mass hierarchy is small.
In this energy range we can neglect the effects of 1-2 mixing and mass
splitting in the first approximation. Then the oscillation
probabilities can be written in the following forms, which are
convenient for discussion of the neutrino mass
hierarchy~\cite{Akhmedov:1998xq}:
\ba P_{ee} & = & 1 - P_A ~, 
\label{eq:probee}\\ P_{\mu e} & = & P_{e\mu}= s_{23}^2 P_A ~, 
\label{eq:probemu}\\ P_{e\tau} & = & c_{23}^2 P_A ~, 
\label{eq:probetau}\\ P_{\mu \mu} & = & 1 - \frac{1}{2} \sin^2 2 
\theta_{23} - s_{23}^4 P_A + \frac{1}{2} \sin^2 2 \theta_{23} \sqrt{1 - 
P_A} \cos \phi_{X}, 
\label{eq:probmumu}\\ P_{\mu \tau} & = & \frac{1}{2} 
\sin^2 2 \theta_{23} - s_{23}^2 c_{23}^2 P_A - \frac{1}{2} \sin^2 2 
\theta_{23} \sqrt{1 - P_A} \cos \phi_{X}, 
\label{eq:probmutau} 
\ea 
where $P_A \equiv |A_{e \tilde{3}}|^2$ and 
\be 
\phi_{X} \equiv arg 
[A_{\tilde{2}\tilde{2}} A_{\tilde{3}\tilde{3}}^*]. 
\nonumber 
\ee 
Here $A_{ij}$ are the amplitudes of
$\tilde{\nu}_i \rightarrow \tilde{\nu}_j$ transitions between the
states of the propagation basis. From these formulas one immediately
sees correlations between different probabilities. The amplitudes can
be represented as
\be 
A_{\tilde{3}\tilde{3}} = |A_{\tilde{3}\tilde{3}}| e^{i \phi_{33}} = 
\sqrt{1 - P_A} e^{i \phi_{33}}, ~~~~ A_{\tilde{2}\tilde{2}} = e^{i 
\phi_{22}} , 
\nonumber 
\ee 
so that $\phi_{X} = \phi_{22} - \phi_{33}$. 

In the case of neutrino propagation in matter of constant density we
obtain explicitly \be \phi_X = \arctan \Big(\!\cos 2\theta_{13}^m
\tan\frac{\phi_{31}^m}{2}\Big) + \frac{V + \Delta}{2}x, \nonumber \ee 
where \be \cos 2 \theta_{13}^m = \frac{\cos 2 \theta_{13} \Delta - 
V}{\sqrt{(\cos 2 \theta_{13} \Delta - V)^2 + \Delta^2 \sin^2 
2\theta_{13} }}, \nonumber \ee \be \phi_{31}^m = x \sqrt{(\cos 2 
\theta_{13} \Delta - V)^2 + \Delta^2 \sin^2 2\theta_{13} } \nonumber \ee 
and \be \Delta \equiv \frac{\Delta m^2_{31}}{2E_\nu}. \nonumber \ee

For matter of constant density   
\be 
P_A = \sin^2 2\theta_{13}^m \sin^2 \frac{\phi_{31}^m}{2}.
\nonumber
\ee

In the limit $\theta_{13} \rightarrow 0$, we obtain 
\be \phi_{X} \approx 
\phi_{32} \equiv \frac{\Delta m_{32}^2 x}{2E}. 
\label{eq:phia} 
\ee 
At the MSW resonance, $\tan^2 \theta_{13}^m = 1$, and the phase
$\phi_{X}$ is again given by (\ref{eq:phia}) (up to a factor
$\cos^2\theta_{13}\approx 1$).  The phase $\phi_X$ approximately
equals $\phi_{32}$ also at energies far above the resonance energy and
coincides with $\phi_{32}$ in vacuum.  Eq.~(\ref{eq:phia}) is
therefore violated only in the resonance region, except at the
resonance point itself. This is a consequence of the smallness of the
mixing angle $\theta_{13}$.

The probabilities for antineutrino channels can be obtained from the
expressions in eqs. (\ref{eq:probee} - \ref{eq:probmutau}) by
substituting $P_A \rightarrow \bar{P}_A$ and ${\phi}_X \rightarrow
\bar{\phi}_X$, where \be \bar{P}_A = P_A (V \rightarrow - V), ~~~~ 
\bar{\phi}_X = \phi_X (V \rightarrow - V). \nonumber \ee In the case of 
normal mass hierarchy (NH), the 1-3 antineutrino mixing in matter is 
suppressed, consequently, $\bar{P}_A \approx 0$ and $\phi_{X}$ nearly 
equals the vacuum phase difference (\ref{eq:phia}).

In the approximation of zero 1-2 splitting, for the inverted neutrino
mass hierarchy (IH) we obtain \be P_A^{IH} = \bar{P}_A^{NH},
~~~ \phi_X^{IH} = - \bar{\phi}_X^{NH}, \nonumber \ee where
$\bar{\phi}_X^{NH} = \phi_X^{NH}(V \rightarrow - V)$ (see the
Appendix).  Therefore \be P_{\alpha \beta}^{IH}
= \bar{P}_{\alpha \beta}^{NH}, ~~~~
\bar{P}_{\alpha \beta}^{IH} = P_{\alpha \beta}^{NH}.  \label{eq:nh-ih} 
\ee The equalities in eq.~(\ref{eq:nh-ih}) receive corrections from 
non-zero 1-2 mass splitting and mixing.

The information about the neutrino mass hierarchy is encoded mainly in
$P_A$ and also in $\phi_X$. If $P_A = 0$ and $\phi_X= \phi_{32}$, the
oscillation probabilities for the IH and NH coincide.

\bfig
\includegraphics[trim=0.5in 0.in 0.5in 0.in, clip, width=3.in]{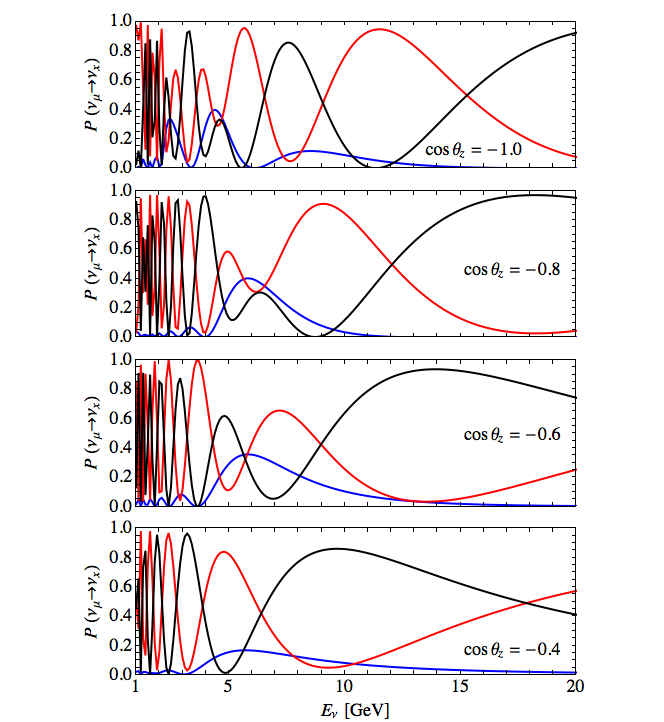}
\includegraphics[trim=0.5in 0.in 0.5in 0.in, clip, width=3.in]{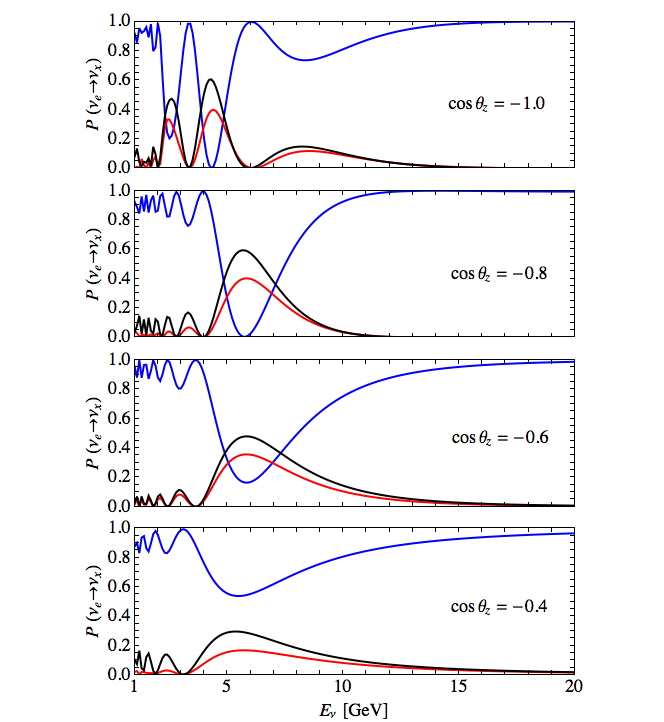}
\caption{
Dependence of the oscillation probabilities in various channels on
neutrino energy for a number of values of the zenith angle. Normal
mass hierarchy is assumed.  Left panel is for the $\nu_\mu
\rightarrow \nu_x$ channels, and right panel is for the $\nu_e \rightarrow 
\nu_x$ channels, where $\nu_x = \nu_e$ (blue lines), $\nu_\mu$ (red lines) 
and $\nu_\tau$ (black lines). We use the values of the neutrino parameters 
defined in the text and $\delta = 0$.}
\label{fig:probab}
\efig

\bfig
\includegraphics[width=2.6in]{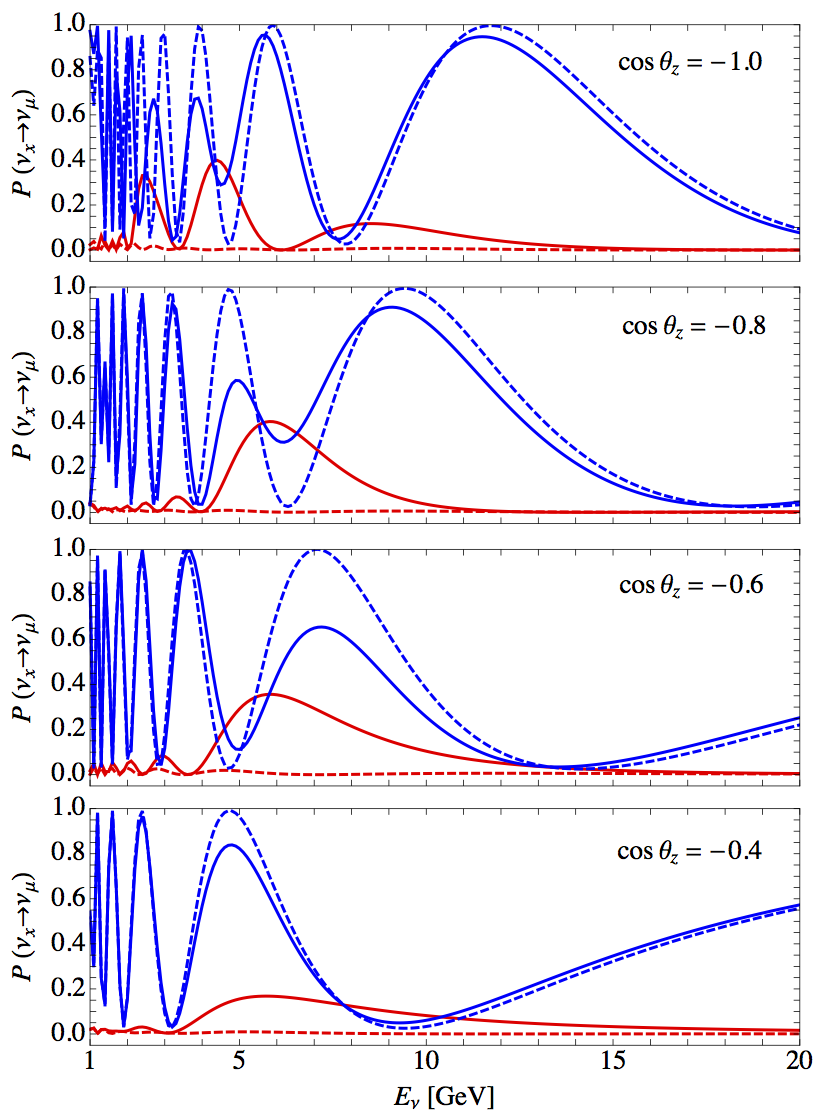}
\hskip 0.35in
\includegraphics[width=2.6in]{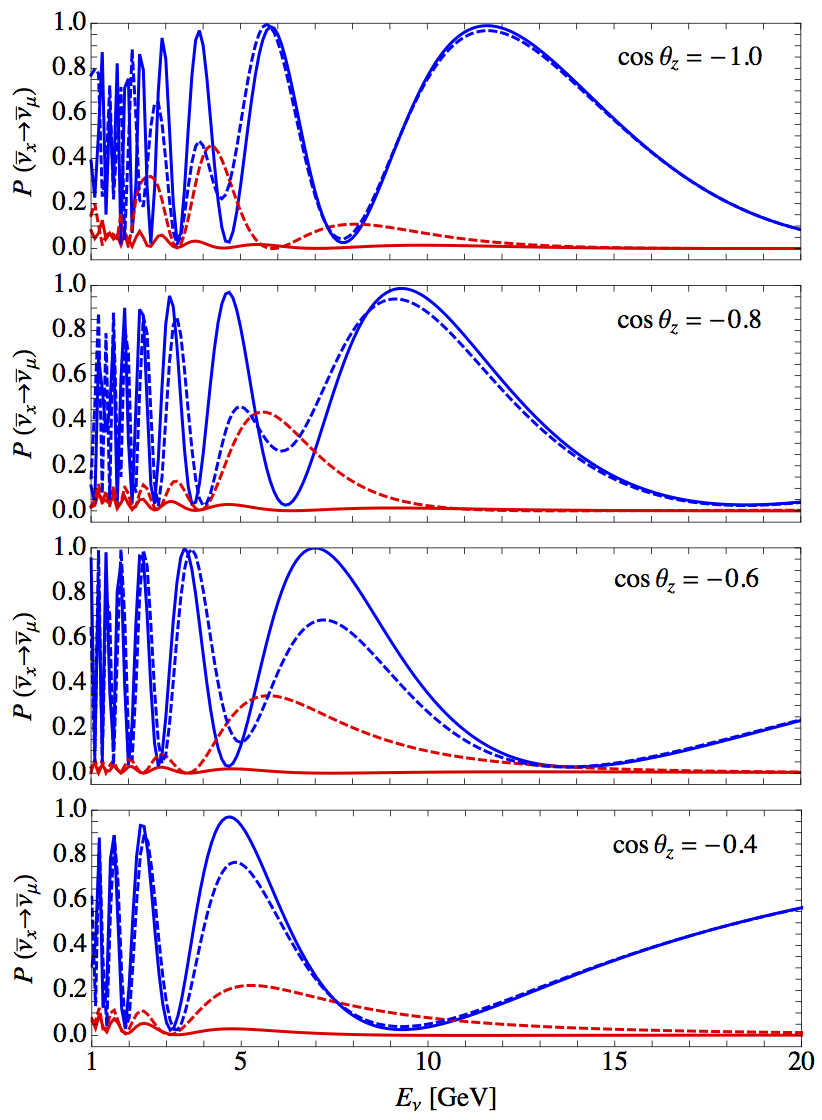}
\caption{
Probabilities of oscillations of neutrinos of various flavors (blue
lines for $\nu_\mu$, red lines for $\nu_e$) to $\nu_\mu$ vs.\ neutrino
energy for a number of values of the zenith angle.  The solid and
dashed lines are for NH and IH, respectively. Left panel is for
neutrinos, and right panel is for antineutrinos. All the parameters
are the same as in Fig.~\ref{fig:probab}.}
\label{fig:probab2}
\efig

Let us consider in more detail eq.~(\ref{eq:probmumu}) for the
probability $P_{\mu \mu}$, which plays a crucial role in our analysis
(a similar analysis can be performed for $P_{\mu \tau}$ as well.) The
first two terms in (\ref{eq:probmumu}) correspond to $\bar{P}$, the
average 2$\nu$ probability with vacuum oscillation depth, which is due
to 2-3 mixing. The probability $P_A$ is an oscillatory function of the
neutrino energy (and the zenith angle), but in the resonance region
the period of oscillations is determined by the oscillation length in
matter $l_{13}^m \approx l_{23}/\sin 2\theta_{13}
\approx (3 - 4) l_{23}$, which is much larger than the 
oscillation length $l_{23}$ responsible for $\phi_X$.  The effects of
the 1-3 mixing (and therefore the matter effects) (i) reduce the
average probability, the third term in (\ref{eq:probmumu}); (ii)
reduce the depths of oscillations by the factor $\sqrt{1 - P_A}$, the
fourth term in (\ref{eq:probmumu}); and (iii) change the oscillation
phase $\phi_{32} \rightarrow \phi_{X}$.  The third and fourth terms
lead to a change of the depths of the oscillations.  These features
are well seen in Fig. \ref{fig:probab}, where we show the dependence
of the oscillation probabilities in various neutrino channels on the
neutrino energy for different values of the zenith angle.  Figure
\ref{fig:probab2} shows the oscillation probabilities for both the 
neutrino and antineutrino channels and for the two mass hierarchies.

We use the values of the neutrino parameters $\Delta m^2_{32} =
2.35 \cdot 10^{-3}$ eV$^2$, $\Delta m^2_{21} = 7.6 \cdot 10^{-5}$
eV$^2$, $\sin^2 \theta_{23} = 0.42$, $\sin^2 \theta_{12} = 0.312$ and
$\sin^2 \theta_{13} = 0.025$, which are close to the current best fit
values~\cite{Fogli:2012ua}, and the PREM density profile of the
Earth \cite{prem} for numerical computations.

The strongest modifications of the oscillation probabilities due to
matter effects are in the resonance region $E_\nu\sim$ (4 - 8) GeV,
where $P_A$ has a peak due to the MSW resonance in the mantle of the
Earth. For the selected values of the oscillation parameters the
maximum of the peak is at $E_\nu = 6.2$ GeV and $\cos \theta_z = -
0.68$.  In addition, at this point the oscillation phase $\phi_{31}^m
= \pi$. In the region $|\cos \theta_z| > 0.83$ and $E_\nu < 7$ GeV
strong modifications of the oscillation probabilities are due to the
parametric enhancement of the oscillations for the neutrino
trajectories crossing the Earth's core, and also due to the MSW
resonance in the core~\cite{our1,our2}.  Notice that $P_{\mu \mu}$,
and consequently the mass hierarchy effects, strongly depend on the
2-3 mixing: the probability $P_A$ enters $P_{\mu \mu}$ with the factor
$s_{23}^4$.

In Fig. \ref{fig:oscillogr} we show the neutrino oscillograms --
curves of equal oscillation probability in the $(E_\nu$ --
$\cos \theta_z)$ plane -- for the $\nu_\mu \rightarrow \nu_x$ and
$\nu_e \rightarrow \nu_x$ channels.  The probabilities increase
monotonically from darker-shaded to lighter-shaded regions.

\bfig 
\includegraphics[trim=0.in 0.in 0.in 0.in, clip, 
width=6.in]{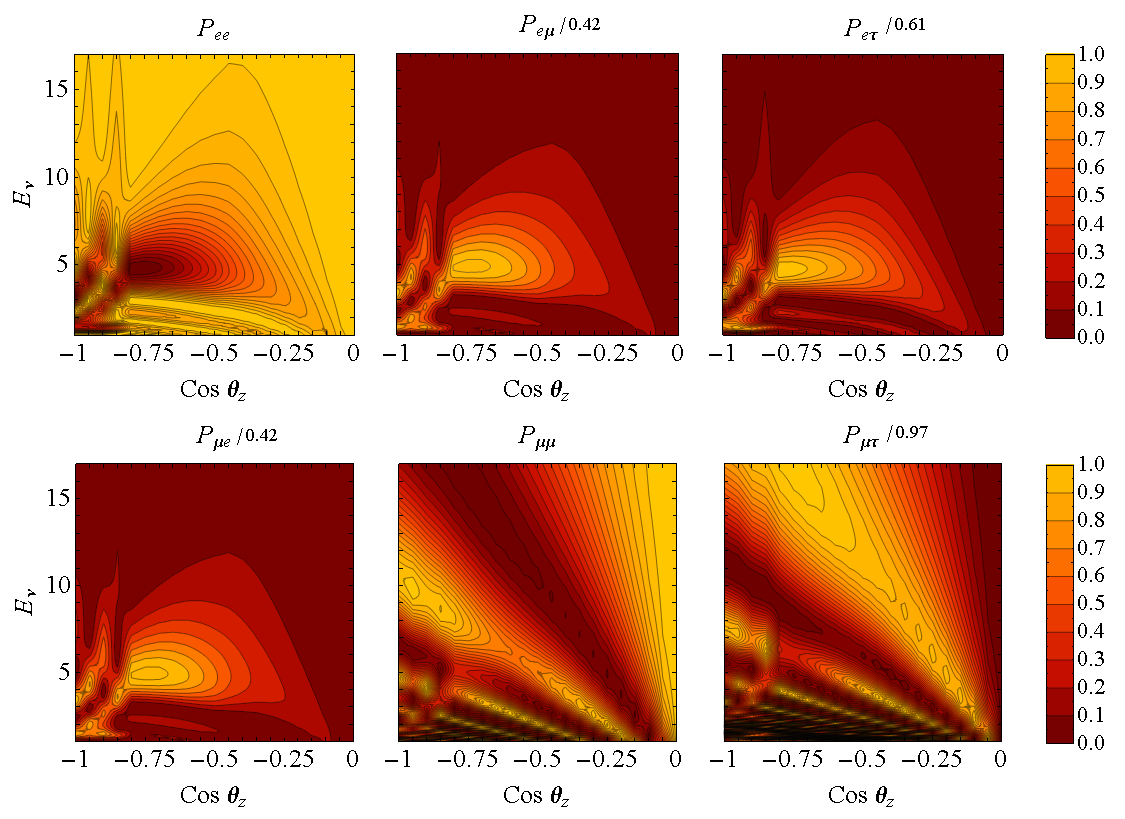} 
\caption{ Neutrino oscillograms of the Earth (lines of equal
  probabilities in the $E_\nu - \cos \theta_z$ plane) for different
  oscillation channels and for the values of the oscillation
  parameters indicated in the text.  Shown are the oscillation
  probabilities normalized by their maximal values in the parameter
  space of the panels: $P_{\alpha, \beta} / P_{\alpha, \beta}^{max} $,
  with $P_{ee}^{max} = P_{\mu \mu}^{max} = 1$. $E_\nu$ is in GeV.
  Normal mass hierarchy is assumed.}
\label{fig:oscillogr}
\efig

\section{Numbers of events in the neutrino energy and zenith angle plane}

We consider first the numbers of events of different types produced by
neutrinos with energies and zenith angles in small bins $\Delta
(E_\nu)$ and $\Delta(\cos \theta_z)$. These would correspond to real
observables if the neutrino energy and zenith angle could be
reconstructed from the measurements with negligible errors.  In the
next section we will study the effects of smearing of these
distributions due to uncertainties of the neutrino energy and zenith
angle reconstructions.

\subsection{Distributions of the $\nu_\mu$-like events}

The $\nu_\mu$-like events (tracks) correspond to interactions $\nu_\mu + 
N \rightarrow \mu + X$, $\bar{\nu}_\mu + N \rightarrow \mu^+ + X$. There 
are also some contributions from $\nu_\tau$ which produce $\tau$ with 
subsequent decay into $\mu$. The number of $\nu_\mu$-like events in the 
$ij$-bin is 
\be N_{ij, \mu}^{\rm NH} = 2 \pi N_A \rho T 
\int_{\Delta_i\cos\theta_z} d\cos\theta_z \int_{\Delta_jE_\nu} dE_\nu~ 
V_{\rm eff} (E_\nu) D_\mu (E_\nu, \theta_z), 
\label{eq:nev} 
\ee 
where $T$ is the exposure time, $N_A$ is the Avogadro's number, $\rho$ is the 
density of ice, $V_{\rm eff}$ is the effective volume of the detector, 
and the number density of events per unit time per target nucleon is 
given by 
\be D_\mu (E_\nu, \theta_z) = \left[\sigma^{CC} 
\left(\Phi_\mu^0 P_{\mu\mu} + \Phi_e^0 P_{e\mu} \right) + {\bar 
\sigma}^{CC} \left({\bar \Phi}_\mu^0 {\bar P}_{\mu\mu} + {\bar \Phi}_e^0 
{\bar P}_{e\mu}\right) \right].  
\label{eq:den} 
\ee 
Here $\Phi_\alpha^0 = \Phi_\alpha^0 (E_\nu,\theta_z)$, are the original 
fluxes of neutrinos $\nu_\alpha$.  We use the effective volume of PINGU 
with 20 strings \cite{cowen} which can be parameterized as 
\be 
\rho V_{\rm eff} (E_\nu) = 14.6 \times [\log(E_\nu/{\rm GeV})]^{1.8} 
~{\rm Mt}. 
\label{eq:rhoV}
\ee 
The volume increases from ~2 Mt at $E_\nu = 2$ GeV to 20 Mt at $E_\nu
= 20$ GeV. (In general $V_{\rm eff}$ depends also on $\theta_z$.)  We
have found $P_A$ and the probabilities $P_{\alpha \beta} = P_{\alpha
\beta} (E_\nu,\theta_z)$ by performing numerical integration of the 
evolution equation for the complete $3\nu-$system. We use the deep 
inelastic cross-sections \ba \sigma^{CC} (E_\nu) = 7.30\cdot 10^{-39} 
(E_\nu/{\rm GeV}) ~{\rm cm}^2, \nonumber \\ {\bar \sigma}^{CC} (E_\nu) = 
3.77\cdot 10^{-39}(E_\nu/{\rm GeV}) ~{\rm cm}^2, \nonumber \ea and we 
take the cross sections for electron and muon neutrinos of the same 
energy to be the same. In our calculations we use the Honda et al.\ 
atmospheric neutrino fluxes \cite{Honda:1995hz}, which were calculated 
for the Kamioka site; however, for neutrino energies above a few GeV 
these should also give good approximations for the fluxes at the South 
Pole.

The fine-binned distribution of events (\ref{eq:nev}) with $\Delta
(\cos \theta_z) = 0.025$ and $\Delta E_\nu = 0.5$ GeV is shown in
Fig.~\ref{fig:event-mu}.  The number of events decreases with $E_\nu$.
The pattern of the event number distribution follows the oscillatory
picture due to the main $\nu_\mu - \nu_\mu$ mode of the oscillations
with a clear distortion in the resonance region. The maxima and minima
are approximately along the lines of equal oscillation phases
$E_\nu \sim \phi_{32}\Delta m^2_{32} |\cos \theta_z| R_E$ (where $R_E$
is the Earth's radius), again with a distortion in the resonance
region $E_\nu = 4 - 10$ GeV.  In the high event density bins the
numbers of event reach $\sim$200, and the total number of events is
about $7\cdot 10^4$.

Introducing the ratios of the fluxes,  
\be
r \equiv \frac{\Phi_\mu^0}{\Phi_e^0},  ~~~~
\bar{r} \equiv \frac{\bar{\Phi}_\mu^0}{\bar{\Phi}_e^0},  
\nonumber
\ee
we can rewrite the expression 
for the density of events (\ref{eq:den}) as 
\be
D_{\mu}^{\rm NH} =  \sigma^{CC} (E_\nu) \Phi_\mu^0  
\left[ \left(P_{\mu\mu}  + \frac{1}{r} P_{e\mu}\right) 
 +
\kappa_\mu \left({\bar P}_{\mu\mu} + \frac{1}{\bar r}  {\bar P}_{e\mu}\right) 
\right], 
\label{eq:mueventsNH}
\ee
where 
\be
\kappa_\mu \equiv
\frac{{\bar \sigma}^{CC} \bar{ \Phi}_\mu^0}
{\sigma^{CC} \Phi_\mu^0}.
\nonumber
\ee
Recall that the ratio $r \equiv \Phi_\mu^0/\Phi_e^0$ depends both on
the neutrino energy and zenith angle, e.g., in the range $E_\nu=(2 -
25)$ GeV and for $\cos\theta_z = -0.8$ the ratio can be roughly
parameterized as $r = 1.2 \cdot (E_\nu/ 1~ {\rm GeV})^{0.65}$.

\bfig
\includegraphics[trim=0.in 0.in 3.in 0.in, clip, width=4.in]{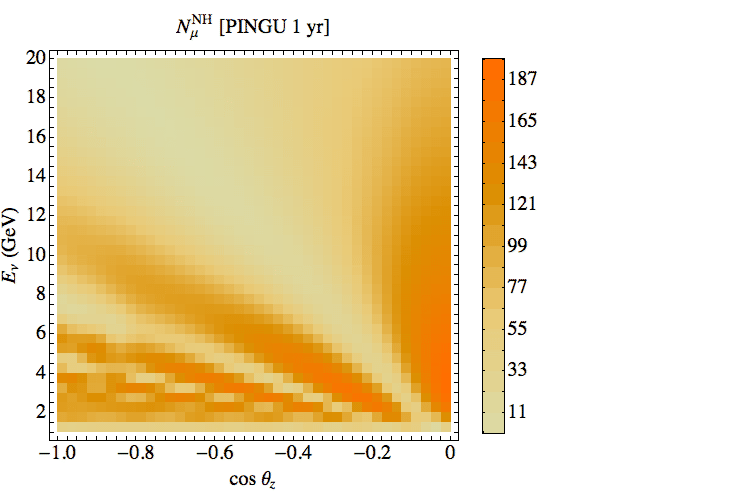}
\caption{
The fine-binned distribution of the number of $\nu_\mu-$like events 
in the $(E_\nu - \cos \theta_z)$ plane that can be collected by the 
PINGU detector during 1 year; NH is assumed.}
\label{fig:event-mu}
\efig

\subsection{Hierarchy asymmetry}

Let us consider the effects of neutrino mass hierarchy on the
distribution of the $\nu_\mu$ events.  Using relations
(\ref{eq:nh-ih}) we can write the density of the events for the
inverted mass hierarchy in terms of the oscillation probabilities for
the normal mass hierarchy as
\be
D_{\mu}^{\rm IH} =  \sigma^{CC} \Phi_\mu^0
\left[ \left(\bar{P}_{\mu\mu}  + \frac{1}{r} \bar{P}_{e\mu}\right)
 +
\kappa_\mu \left({P}_{\mu\mu} + \frac{1}{\bar r}  {P}_{e\mu}\right)
\right]. 
\label{eq:mueventsIH}
\ee
Then the difference of the numbers of events  for the inverted
and normal mass hierarchies equals 
\be
N_{ij, \mu }^{\rm IH}  -  N_{ij, \mu }^{\rm NH}  =  {2\pi N_A \rho T}
\int_{\Delta_i\cos\theta_z} d\cos\theta_z \int_{\Delta_jE_\nu} dE_\nu
V_{\rm eff} (D_{\mu }^{\rm IH}  -  D_{\mu }^{\rm NH}), 
\nonumber
\ee 
where
\be 
D_{\mu }^{\rm IH}  -  D_{\mu }^{\rm NH} =   
\sigma^{CC} \Phi_\mu^0 \left[ 
(1 - \kappa_\mu) \left(\bar{P}_{\mu\mu} - P_{\mu\mu} \right) 
+     
\frac{1}{r}  \left(1 - \kappa_e \right)         
\left(\bar{P}_{e\mu} - P_{e\mu} \right)
\right],  
\label{eq:diff}
\ee
and 
\be
\kappa_e \equiv 
\frac{{\bar \sigma}^{CC} \bar{ \Phi}_e^0}
{\sigma^{CC} \Phi_e^0} = \kappa_\mu \frac{r}{\bar{r}}~. 
\nonumber
\ee
In the approximation $\bar{P}_A \approx 0$, which is justified if the 
true hierarchy is the normal one, we obtain $\bar{P}_{e \mu} \approx 0$ and  
\be
\bar{P}_{\mu \mu}  \approx  1 - \frac{1}{2} \sin^2 2 \theta_{23}
\left[1 - \cos \phi_{32} \right],  
\nonumber
\ee
where we have taken $\bar{\phi}_{X} \approx \phi_{32}$. 
Consequently, 
\ba
\bar{P}_{\mu \mu} - P_{\mu \mu} &  \approx &   
\frac{1}{2} \sin^2 2 \theta_{23}
\left[ \cos \phi_{32} - \sqrt{1 - P_A} \cos \phi_{X}\right] + s_{23}^4 P_A ~,
\nonumber\\
\bar{P}_{e \mu} - P_{e \mu} &  \approx & - s_{23}^2 P_A . 
\label{eq:difemu}
\ea
In this approximation 
\ba
D_{\mu}^{\rm NH} & \approx & \sigma^{CC} \Phi_\mu^0
\left[\left(1- \frac{1}{2} \sin^2 2\theta_{23} \right) 
(1 + \kappa_\mu) -  s_{23}^2 \left(s_{23}^2 - \frac{1}{r} \right)P_A  + 
\right.
\nonumber\\
&  + & 
\left.\frac{1}{2} \sin^2 2 \theta_{23}
\left(\kappa_\mu \cos \phi_{32} + \sqrt{1 - P_A} \cos \phi_{X} \right)
\right]. 
\nonumber
\ea
The sensitivity of this quantity to the neutrino mass hierarchy is due to 
$P_A \neq 0$, and so the highest sensitivity is expected in the kinematic 
region where $P_A$ is relatively large.

For the difference of the numbers of events for the two hierarchies 
(\ref{eq:diff}) we obtain  
\ba
D_{\mu}^{\rm IH} -  D_{\mu}^{\rm NH} 
& \approx &  \sigma^{CC} \Phi_\mu^0 
\left\{ \frac{1}{2} \sin^2 2 \theta_{23} (1 - \kappa_\mu)
\left(\cos \phi_{32} - \sqrt{1 - P_A} \cos \phi_{X} \right) + 
\right.
\nonumber\\
& + & 
\left. s_{23}^2 \left[ (1 - \kappa_\mu) s_{23}^2 - 
\left(\frac{1}{r} - \frac{\kappa_\mu}{\bar{r}}\right) \right] P_A
\right\}.
\nonumber
\ea

Let us introduce the N-I hierarchy asymmetry for the $ij$-bin  in the 
$(E_\nu - \cos \theta_z)$ plane as 
\be
A^{N-I}_{\mu, ij}  \equiv \frac{N^{IH}_{\mu, ij} 
- N^{NH}_{\mu,ij}}{\sqrt{N^{NH}_{\mu, ij}}}. 
\label{eq:asym}
\ee
The moduli of the asymmetries (\ref{eq:asym}) are the measure of
statistical significance of the difference of the number of events for
the normal and inverted mass hierarchies: $S_{ij} = |A_{ij}|$.

Let us consider the condition $N_{ij, \mu}^{\rm IH}=N_{ij, \mu}^{\rm NH}$ 
which gives the borders of the regions in the $(E_\nu - \theta_z)$ plane   
where the difference of the numbers of events has definite sign.
It coincides approximately with the condition  $D_{\mu}^{\rm IH}=
D_{\mu}^{\rm NH}$.  The latter determines the lines of zero N-I hierarchy 
asymmetry. Using eq.~(\ref{eq:diff}) and approximate expressions in 
(\ref{eq:difemu}) we find from this condition 
\be
\cos \phi_{32} -  \sqrt{1 - P_A} \cos \phi_X = \frac{P_A}{2 c_{23}^2}
\left[ \frac{1}{r} \cdot \frac{1 -  \kappa_e}{1 - \kappa_\mu} - s_{23}^2 \right]. 
\label{eq:rel2}
\ee
The phases  $\phi_{32}$ and  $\phi_{X}$  are functions of $E_\nu$ and 
$\theta_z$.  Since $\cos \phi_{32}$ varies with  
$(\cos \theta_z/E_\nu)$ much faster than $r(E_\nu,  \theta_z)$, it is this 
periodic function that determines the lines of zero hierarchy asymmetry. 
Our calculations show that eq.~(\ref{eq:rel2}) determines the zero 
asymmetry lines rather well.

\bfig
\includegraphics[trim=0.in 0.in 0.in 0.in, clip, width=4.in]{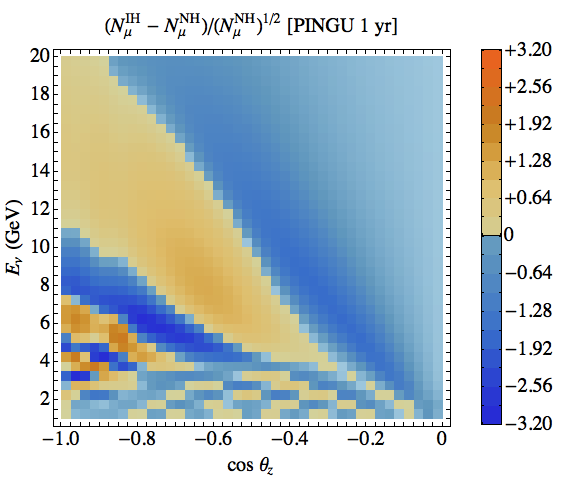}
\caption{
The hierarchy asymmetry of $\nu_\mu$ events in the $E_\nu
- \cos \theta_z$ plane.  The absolute value of the asymmetry in a
given bin determines the statistical significance of the difference of
the numbers of events for the inverted and normal mass hierarchies.}
\label{fig:asym-mu}
\efig

In Fig.~\ref{fig:asym-mu} we show the values of the hierarchy
asymmetry in the $E_\nu - \cos \theta_z$ plane.  The maximal asymmetry
is achieved in the low energy parts of the domains bounded by the
lines of zero asymmetry (\ref{eq:rel2}). The large asymmetry is in the
strips along the constant phase lines in the energy interval
$E_\nu \approx (4 - 12)$ GeV, where these lines are distorted by
matter effects.  The asymmetry changes the sign with changing zenith
angle. For instance, in the energy range (7 - 11) GeV there are three
distinct zenith angle intervals with the asymmetry sign being the same
within each interval.  The number of such intervals increases with
decreasing energy. Therefore, due to the averaging, the region of high
sensitivity to hierarchy will shift to higher energies if the zenith
angle reconstruction becomes worse.  Furthermore, the regions of high
significance of the hierarchy determination overlap substantially with
the regions of small numbers of events (see Fig.~\ref{fig:event-mu}).
This means that the significance is enhanced due to the smallness of
the denominator in eq.~(\ref{eq:asym}), and it would be diluted by
combining a given bin with bins which have higher statistics but
smaller significance.

There is an important background to the $\nu_\mu$ events which comes
from $\nu_{\tau}$ interactions.

\subsection{$\nu_{\tau}$ events}

The $\nu_{\tau}$ flux appears at the detector due to $\nu_\mu -
\nu_\tau$ oscillations. In Fig.~\ref{fig:event-tau} 
we show the distribution of the $\nu_\tau$ CC events ($\nu_\tau + N 
\rightarrow \tau +X$) in the $E_\nu - \cos \theta_z$ plane. The figure is 
a kind of inversion of Fig.~\ref{fig:event-mu}, with maxima
substituted by minima and {\it vice versa}. The number of events is,
however, smaller than the number of $\nu_\mu$ events due to the
smaller cross-section near the threshold. Notice that since $\nu_\tau$
(as well as $\nu_l$) from the sequential $\tau$ decays
($\tau \rightarrow \nu_\tau + X$, $\tau \rightarrow \nu_\tau + l
+ \nu_l$) is not detected, the energy of the original $\nu_\tau$
cannot be reconstructed.

\bfig
\includegraphics[trim=0.in 0.in 2.2in 0.in, clip, width=4.in]{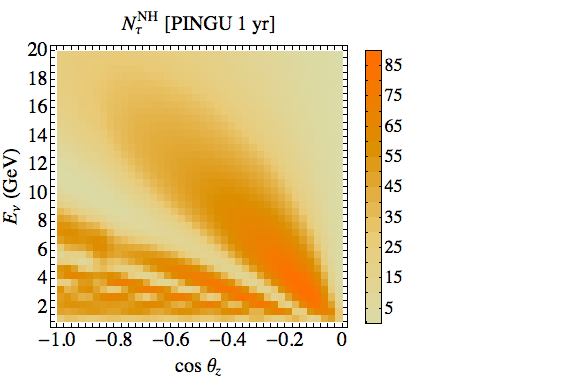}
\caption{
Same as in Fig. \ref{fig:event-mu}, but for $\nu_\tau$ CC interactions.}
\label{fig:event-tau}
\efig

The $\nu_{\tau}$ interactions
\be
\nu_{\tau} + N \rightarrow \tau
+  h \rightarrow \mu + \nu + \nu + h
\label{nutau}
\ee
will contribute to the main sample of $\nu_\mu$ events with a muon and
a hadronic cascade in the final state.  However, the number of these
events is relatively small, and in addition these events have certain
features which can be used to discriminate them from the true
$\nu_\mu$ events.

Indeed, on average the two neutrinos which appear in the process
(\ref{nutau}) will take about 1/3 of the energy of the initial
neutrino.  Therefore, for a given observed total energy $E_\mu + E_h$,
the energy $E_0(\nu_\tau)$ of the original neutrino in the process
(\ref{nutau}) should be about 1.5 times larger than the energy
$E_0(\nu_\mu)$ of the true $\nu_\mu$ event:
$E_0(\nu_\tau)/E_0(\nu_\mu)\approx 1.5$. If only $E_\mu$ is used to
reconstruct the energy of the original neutrino the rescaling
coefficient is $E_0(\nu_\tau)/E_0(\nu_\mu) = 2.5$. On average we can
take a factor of 2 for our estimates. Furthermore, the branching ratio
of tau decay into muon is $B_\mu = 0.17$. Consequently, the number of
$\nu_\mu$ events due to the reaction chain (\ref{nutau}) is suppressed
with respect to that of the true CC $\nu_{\mu}$ events of the same
energy by a factor
\be 
B_\mu \left( 
\frac{E_0(\nu_\tau)}{E_0(\nu_\mu)} \right)^{-1.3} 
\frac{\sigma_\tau}{\sigma_\mu} 
\label{eq:factor}
\ee 
(provided that the initial $\nu_\mu$ and $\nu_{\tau}$ fluxes are
equal). The power of the second factor follows from the energy
dependences of the neutrino flux $(\propto E^{-3})$, the cross-section
($\propto E$) and the effective volume (which we take to be $\propto
E^{0.7}$ here; note that in the energy range $E_\nu = 10-35$ GeV such
a simple power law approximates eq.~(\ref{eq:rhoV}) within 5\% error).
The last factor takes into account the threshold effect for the $\tau$
production.  For the energy rescaling factor of 1.5 - 2 we obtain a
suppression factor for the number of $\nu_{\tau}$-induced $\nu_\mu$
events to be $0.05 - 0.08$. Due to the missing energy and momentum
taken by the two neutrinos in the final state of reaction
(\ref{nutau}), the smearing effects in the energy and angle of the
original neutrino around the average values will be stronger than for
the true $\nu_\mu$ events. Therefore, the observed energy $E_\mu +
E_h$ will have a bigger spread for $\nu_\tau$-induced $\nu_\mu$
events. Furthermore, they will be characterized by a larger average
angle between the momenta of the original neutrino and the muon.

There are other properties of reaction (\ref{nutau}) which can be
utilized to disentangle it from the CC $\nu_\mu$ detection
reaction. In particular, correlations between $E_\mu$ and $E_h$ are
different for these two cases. Furthermore, one can select $E_\nu
- \theta_z$ regions in which the $\nu_\mu-\nu_\tau$ transition
probability, and consequently the $\nu_{\tau}$ flux, are
suppressed. Such regions can be readily found with the help of
Fig.~\ref{fig:event-tau}.  The corresponding restriction of the $E_\nu
- \theta_z$ parameter space will, of course, result in a loss of the
overall statistics, but would provide us with cleaner events. The
resulting statistics loss should be affordable because of the
extremely high overall statistics in multi-megaton detectors.

In principle, one could also sum up the $\nu_\mu$ and $\nu_\tau$
events and consider them in the $\theta_\mu - (E_\mu + E_h)$
plane. However, to determine whether or not a useful information can
be extracted from such data would require an additional analysis which
is outside the scope of the present paper.

For the above reasons, in what follows we do not explicitly consider
the contributions of the $\nu_\tau \rightarrow \tau \rightarrow \mu$
events and simply treat them as a $5\%$ systematic error.

\subsection{Cascade events and the mass hierarchy}
Following the IceCube terminology, we will call the events in which
the muon track is not identified as cascade events.  There are several
different contributions to the cascade events, including even
$\nu_\mu$ events with faint muon tracks which can not be identified.
All $\nu_\tau$-induced CC events, except those when the tau decays
into a muon, contribute to the cascade events.

For the cascade events $\nu_e + N \rightarrow e + X$ 
and  $\bar{\nu}_e + N \rightarrow e^+  + X$ we have
\be
N_{ij, e}^{\rm NH}  = 2 \pi N_A \rho T 
\int_{\Delta_i\cos\theta_z} d\cos\theta_z \int_{\Delta_jE_\nu} dE_\nu
V_{\rm eff}(E_\nu) D_e (E_\nu, \cos\theta_z), 
\nonumber
\ee
where
\be
D_e (E, \cos\theta_z) =  
\sigma^{CC} \Phi_e^0
\left[ \left(P_{ee}  + r P_{\mu e}\right)
 +
\kappa_e \left({\bar P}_{ee} + \bar{r} {\bar P}_{\mu e}\right)
\right]. 
\nonumber
\ee
In terms of the probability $P_A$ the number density of events can be 
written as 
\be
D_{e}^{NH} = \sigma^{CC} \Phi_e^0
\left\{ 1 + P_A (r s_{23}^2 - 1) +   
\kappa_e [1 +  \bar{P}_A (\bar{r} s_{23}^2 - 1)]
\right\}. 
\nonumber
\ee
The event distribution is shown in Fig.~\ref{fig:event-e}.  Notice
that here the distribution is weakly affected by the oscillations due
to a substantial screening effect: the oscillatory part of the number
of events contains terms proportional to $r s^2_{23} - 1$ and $\bar{r}
s^2_{23} - 1$ which are nearly zero at low energies.

\bfig
\includegraphics[trim=0.in 0.in 2.in 0.in, clip, width=4.in]{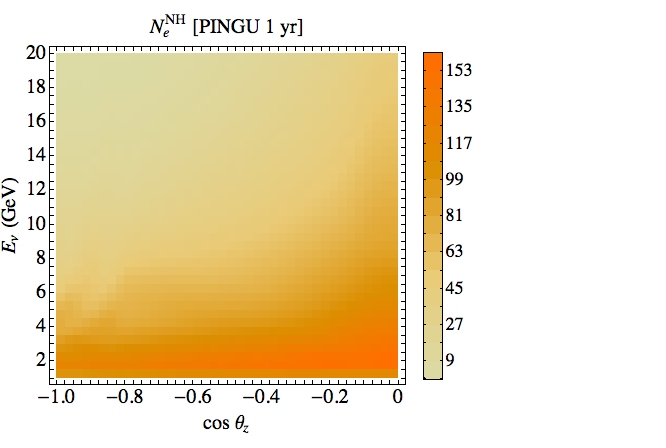}
\caption{
Same as in Fig. \ref{fig:event-mu} but for $\nu_e$ CC interactions; NH
is assumed.}
\label{fig:event-e}
\efig

The difference of the number densities of events for the inverted and 
normal mass hierarchies is 
\ba
D_{e}^{\rm IH} -   D_{e}^{\rm NH} = 
\sigma^{CC}\Phi_e^0
~(\bar{P}_A - P_A)  
\left[ (r s_{23}^2 - 1) - \kappa_e (\bar{r} s_{23}^2 - 1) \right].  
\label{eq:easym}
\ea
The expression in the square brackets here can be rewritten as 
\be
r s_{23}^2 (1 - \kappa_\mu) - (1 - \kappa_e) = 
(1 - \kappa_\mu)\left[r s_{23}^2 - \frac{1 - \kappa_e}{1 - \kappa_\mu}\right].   
\label{eq:suppressions}
\ee
Thus, there is a double suppression of the difference of numbers of
events: (i) due to the neutrino-antineutrino factor $(1 - \kappa_\mu)$
related to the presence of both the neutrino and antineutrino fluxes,
and (ii) due to the flavor screening (the last term in
eq. (\ref{eq:suppressions})) related to the presence of both $\nu_e$
and $\nu_\mu$ in the the original atmospheric neutrino
flux~\cite{Akhmedov:1998xq}, more precisely, due to the ratio of these
fluxes being close to 1/2 at low energies.  The difference $D_{e}^{\rm
IH} - D_{e}^{\rm NH}$ can be further suppressed by the smallness of
$P_A$ (or of the difference of the neutrino and antineutrino
probabilities $\bar{P}_A - P_A$).

The numbers of events for IH and NH are equal, $N_{ij, e}^{\rm IH} =  
N_{ij, e}^{\rm NH}$, in the bins for which both sides of  
eq.~(\ref{eq:suppressions}) vanish. From this we obtain 
\be 
s_{23}^2(r - \bar{r}\kappa_e) =  (1 - \kappa_e). 
\nonumber
\ee
For  $r \approx \bar{r}$ it gives 
\be
r (E_\nu, \theta_z ) = \frac{1}{s_{23}^2}. 
\nonumber
\ee
So, in this approximation we have only one line of zero asymmetry. 

The hierarchy asymmetry in the $\nu_e$ CC events is shown in
Fig.~\ref{fig:asym-e}. Maximal asymmetry is in the resonance region
$E_\nu = (4 - 8)$ GeV where it has essentially the same sign, so that
the suppression due to averaging is absent. Unfortunately, other
contributions to the cascade events have different pattern in the
$E_\nu - \theta_z$ plane. The number of cascades induced by $\nu_e$
quickly decreases with the increase of the neutrino energy.

\bfig
\includegraphics[trim=0.in 0.in 1.75in 0.in, clip,width=4.in]{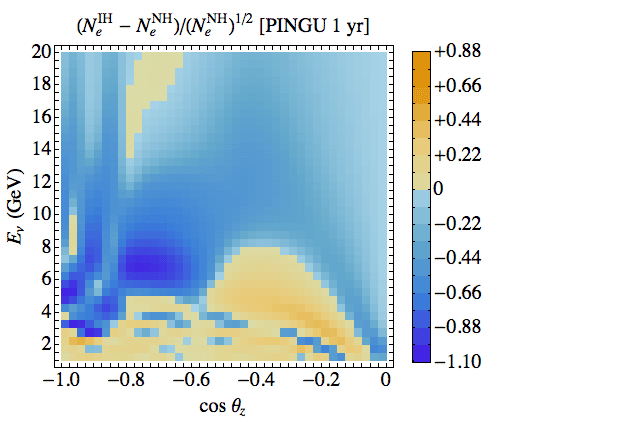}
\caption{Same as in Fig. \ref{fig:asym-mu} but for $\nu_e$ induced 
events.}
\label{fig:asym-e}
\efig

The dominant contribution to the cascade events comes from the
$\nu_\tau$ flux which appears at the detector due to the oscillations:
\be
D_\tau^{NH}  = \sigma_\tau^{CC} (E_\nu) 
\Phi_{\mu}^0\left\{\frac{1}{2} \sin^2 2 \theta_{23} 
[1 - \sqrt{1 - P_A} \cos \phi_{X}]  
- c_{23}^2 P_A (s_{23}^2 - 1/r) \right\}.  
\nonumber
\ee
For the $\nu_\tau$ events the screening due to the flavor composition
of the original neutrino flux is absent.

The neutral current interactions of all neutrino species contribute to
the total numbers of the cascade events but do not affect the NH-IH
differences of events.  This reduces the hierarchy asymmetry:
$$
A_{cascades} = \frac{N^{IH}_{e + \tau} - 
N^{NH}_{e + \tau}}{\sqrt{N^{NH}_{e + \tau} + N_{NC}}}.  
$$

As we will see, analysing only $\nu_\mu$ events will be sufficient to
establish the neutrino mass hierarchy. We therefore do not include
cascade events in our discussion. Clearly, cascade events can give
additional information, but taking them into account would require a
more sophisticated analysis. Notice that possible identification of
$\tau$ events in a large liquid argon detector and its consequences
have been discussed in \cite{Conrad}; should such an identification
turn out to be possible also in PINGU, it would increase PINGU's
sensitivity to the neutrino parameters.

\subsection{Effects of deviation of the 2-3 mixing from the maximal one}

We describe deviation of the 2-3 mixing from the maximal one by  
\be
d_{23} \equiv \frac{1}{2} - s^2_{23}. 
\nonumber
\ee
From eqs.~(\ref{eq:mueventsNH}), (\ref{eq:probemu})  and~(\ref{eq:probmumu})
we find 
\ba
D_{\mu}^{NH} (\theta_{23})  & - & D_{\mu}^{NH} (\pi/4)    
 \approx  \sigma^{CC} \Phi_\mu^0
\left\{ 2 d_{23}^2  
\left[ 1 - \sqrt{1 - P_A} \cos \phi_{X} 
\kappa_\mu (1 - \cos \phi_{32}) \right] 
\right.
\nonumber\\
& + & 
\left. d_{23}  \left(1 - \frac{1}{r}  - d_{23}\right)P_A
\right\}.  
\label{eq:23sesitivity}
\ea 
Notice that both terms in (\ref{eq:23sesitivity}) are positive
for $\theta_{23} < \pi/4$. 

In Fig. \ref{fig:event23mu} we plot statistical significance of the
determination of a deviation of the 2-3 mixing from the maximal one.
Here again high significance regions coincide with the regions of low
density of events (compare Figs. \ref{fig:event23mu} and
\ref{fig:event-mu}). This means that an integration over large 
$E_\nu - \cos \theta_z$ domains would lead to a dilution of the  
significance.

\bfig
\includegraphics[trim=0.in 0.in 3.in 0.in, clip, width=4.in]{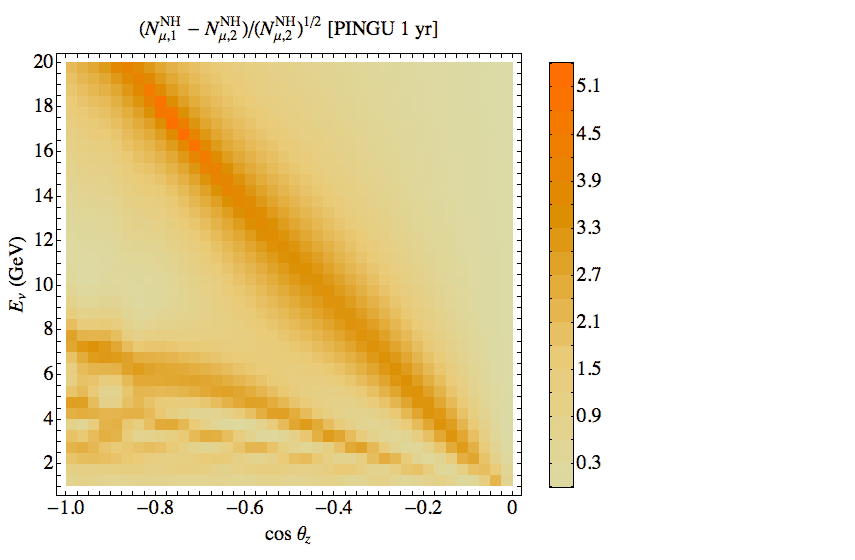}
\caption{
Statistical significance of the determination of the deviation of the
2-3 mixing from the maximal one. The difference of the tracking events
for two values of the 2-3 mixing: $\sin^2 \theta_{23} = 0.5$
($N_{\mu,1}^{NH}$) and 0.42 ($N_{\mu,2}^{NH}$).}
\label{fig:event23mu}
\efig

\subsection{CP-violation effects}

We can write the oscillation probabilities in the presence of a
Dirac-type CP-phase $\delta$ as
\be
P_{\alpha\beta}=P_{\alpha\beta}^0+P_{\alpha\beta}^\delta\,.
\label{eq:genProb}
\ee
Here $P_{\alpha\beta}^0$ and $P_{\alpha\beta}^\delta$ are the
$\delta$-independent and $\delta$-dependent parts of the oscillation
probability $P_{\alpha\beta}$, respectively (note that
$P_{\alpha\beta}^0 \ne P_{\alpha\beta}^{\delta=0}$).  In a matter with
symmetric density profile one has $P_{\beta\alpha}=
P_{\alpha\beta}(\delta\to -\delta)$.

Now the 1-2 mass splitting and mixing should be taken into account.
The state $\tilde{\nu}_2$ does not decouple and the oscillation
probabilities depend on the matrix elements $A_{e \tilde{2}}$,
$A_{\tilde{2}\tilde{3}}$ of the evolution matrix $A$ in the
propagation basis. Since the amplitude $A_{\tilde{2}\tilde{3}}$ is
doubly suppressed (by small $\Delta m_{21}^2/ \Delta m_{31}^2$ and
$s_{13}$), the terms that are quadratic in $A_{\tilde{2}\tilde{3}}$
can be neglected. We have then \cite{our3}:
\ba
P_{ee}^\delta & = & 0, 
\nonumber\\
P_{e\mu}^\delta &  = & \sin 2\theta_{23}|A_{e\tilde{2}}A_{e\tilde{3}}
|\cos(\phi+\delta)\,,
\label{eq:Pemudelta}\\ 
P_{\mu\mu}^\delta &  = & -\sin 
2\theta_{23}\cos\delta\left\{
|A_{e\tilde{2}}^{}A_{e\tilde{3}}^{}|\cos\phi
+\cos 2\theta_{23}{\rm Re}[A_{\tilde{2}\tilde{3}}^*(A_{\tilde{3}\tilde{3}}^{}-
A_{\tilde{2}\tilde{2}}^{})]\right\}\,,
\label{eq:Pmumudelta}
\ea
where $\phi\equiv {\rm arg}(A_{e\tilde{2}}^*A_{e\tilde{3}}^{})$.  We
will also use the notation $|A_{e\tilde{2}}^{}|\equiv \sqrt{P_S}$,
$|A_{e\tilde{3}}^{}|\equiv \sqrt{P_A}$ (where $P_A$ now depends also
on the parameters of the 1-2 sector).

Note that the last term in the curly brackets in
eq.~(\ref{eq:Pmumudelta}) is small if the 2-3 mixing is sufficiently
close to the maximal one, and in addition the amplitude
$A_{\tilde{2}\tilde{3}}$ is small. We shall therefore use for our
estimates the approximation
\be
\cos 2\theta_{23}{\rm Re}[A_{\tilde{2}\tilde{3}}^*(A_{\tilde{3}\tilde{3}}^{}-
A_{\tilde{2}\tilde{2}}^{})]\approx 0\,.
\label{eq:approx1}
\ee 
The $\delta$-dependent part of the expression for the number density
of the $\mu$-like events is
\be
D_\mu^\delta\equiv 
\sigma^{CC} \Phi_\mu^0 \left[\left(P_{\mu\mu}^\delta+\frac{1}{r} 
P_{e\mu}^\delta\right) 
+\kappa_\mu \left(\bar{P}_{\mu\mu}^\delta+\frac{1}{\bar{r}} 
\bar{P}_{e\mu}^\delta\right)\right].
\label{eq:Nmudelta1}
\ee
Next, we notice that in the case of normal hierarchy  
one has $\bar{P}_S\approx 0$, $\bar{P}_A\approx 0$, and in the approximation 
(\ref{eq:approx1}) we therefore have $\bar{P}_{e\mu}^\delta\approx 
\bar{P}_{\mu e}^\delta\approx 0$, $\bar{P}_{\mu\mu}^\delta\approx 0$. From 
eqs.~(\ref{eq:Pemudelta}) and (\ref{eq:Pmumudelta}) we find   
\be
D_\mu^\delta - D_\mu^{\delta=0} ~=  \sigma^{CC} 
\Phi_\mu^0 \sin 
2\theta_{23}\sqrt{P_A 
P_S}\left[\frac{r - 1}{r}\cos\phi(1 - \cos\delta) - \frac{1}{r}\sin\phi 
\sin\delta\right].
\label{eq:Nmudelta2}
\ee
This, in particular, means that the difference $N_\mu^\delta
-N_\mu^{\delta=0}$ should nearly vanish whenever $P_S=0$ or $P_A=0$,
i.e. along the so-called solar and atmospheric ``magic''
lines \cite{Barger:2001yr, Huber:2003ak,Smirnov:2006sm,our3}. The
vanishing of the difference $N_\mu^\delta - N_\mu^{\delta=0}$ is,
however, not exact, as it relies on the approximation
(\ref{eq:approx1}). In Figs.~\ref{fig:event-cp05} -
\ref{fig:event-cp025} we show statistical significance of measurements of 
the CP-phase. The strongest effect is at low energies: $E \sim 3 - 5$
GeV.  Notice that with increasing $\delta$ the size of the asymmetry
increases, but the regions of different signs of the asymmetry do not
change. This is in agreement with eq.~(\ref{eq:Nmudelta2}). Indeed, in
this equation the dependences of the right-hand side on $\phi$ and
$\delta$ effectively factorize, because in most of the parameter space
either the first or the second term dominates.  For the $\nu_e$-like
events we obtain similarly
\be
D_e^\delta \equiv \sigma^{CC} \Phi_e^0 
\left[\left(P_{ee}^\delta+r P_{\mu e}^\delta\right) 
+\kappa_e \left(\bar{P}_{ee}^\delta+\bar{r}\bar{P}_{\mu e}^\delta\right)\right]
\approx 
\sigma^{CC}
\Phi_e^0 \,r\, \sqrt{P_A P_S} \cos(\phi-\delta)\,. 
\label{eq:Nedelta1}
\ee
Here we have taken into account that $P_{ee}$ and $\bar{P}_{ee}$ are
$\delta$-independent and that $\bar{P}_{\mu e}$ is strongly suppressed
in matter. For the difference of the densities of $\nu_e$-like events
we then find
\be
D_e^\delta - D_e^{\delta=0}  = \sigma^{CC} 
\Phi_e^0 r \sin 2\theta_{23}\sqrt{P_A 
P_S}\left[\cos\phi(\cos\delta-1)+\sin\phi \sin\delta\right].
\label{eq:Nedelta2}
\ee

The lines of zero $P_A$ and $P_S$ (the atmospheric and solar ``magic''
lines) form a grid, which leads to a domain structure in the $(E_\nu -
\cos\theta_z)$ plane. For maximal 2-3 mixing the domain structure of 
the distribution of the events becomes sharper (Fig.~\ref{fig:event-cp052}). 
As discussed in ref.~\cite{our3}, in the full $3\nu$ framework there is no 
crossing of the ``magic'' lines determined by the solar, atmospheric and an 
additional phase condition, and instead there is a smooth transition of lines 
of different types into each other.

\bfig
\includegraphics[trim=0.in 0.in 0.in 0.in, clip, width=4.in]{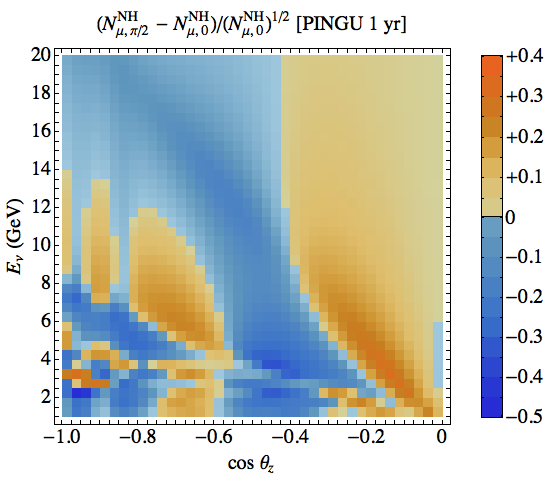}
\caption{
The difference of the numbers of events for $\delta =\pi/2$ and
$\delta = 0$. Statistical significance of measurements of the
CP-phase.}
\label{fig:event-cp05}
\efig

\bfig
\includegraphics[trim=0.in 0.in 3.1in 0.in, clip, width=4.in]{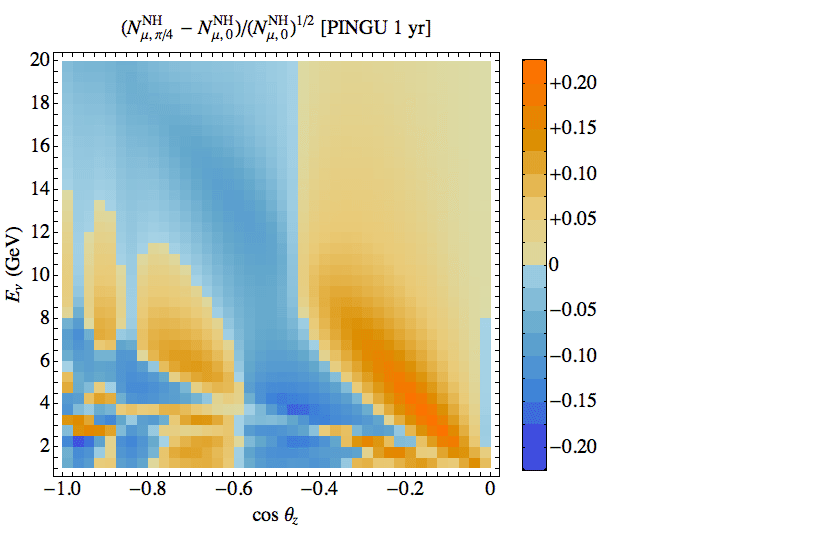}
\caption{
Same as in Fig. \ref{fig:event-cp05} but for the phases $\delta
=\pi/4$ and $\delta = 0$ .}
\label{fig:event-cp025}
\efig

\bfig
\includegraphics[trim=0.in 0.in 3.1in 0.in, clip, width=4.in]{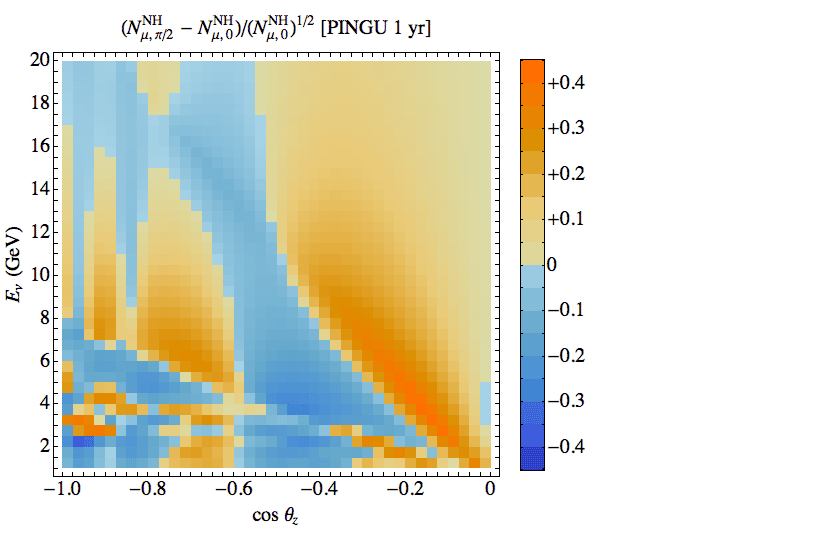}
\caption{
Same as in Fig. \ref{fig:event-cp05} but for phases $\delta =\pi/2$
and $\delta = 0$ and $\sin^2 \theta_{23} = 0.5$.}
\label{fig:event-cp052}
\efig

\section{Determination of the neutrino mass hierarchy}

The fine-binned event distributions computed in the previous section
allow us to identify the regions of high sensitivity to the neutrino
mass hierarchy as well as to other neutrino parameters. They show the
lines of zero asymmetry which separate the regions of same sign
asymmetry.  These distributions also allow one to identify the regions
of high and low degeneracy between different parameters. By using them
one can select the regions of integration over the neutrino energy and
zenith angle in such a way that the hierarchy asymmetry is enhanced
while the effects of the uncertainties due to the incomplete knowledge
of the other neutrino parameters are canceled or
suppressed. Obviously, one should avoid integrating over regions with
different signs of the hierarchy asymmetry.

\subsection{Summation of events over different bins}

Since one needs to integrate over certain regions of the neutrino
energy and zenith angle, the significances of determination of the
neutrino mass hierarchy and of other neutrino parameters will be
modified in comparison to those found for small individual bins.  The
combined statistical significance resulting from a summation over $n$
bins is given by
\be
S_n = \sum_{i = 1}^n  S_i \sqrt{\frac{N_i^{NH}}{ \sum_{k = 1}^n N_k^{NH}}}.  
\nonumber
\ee
If the individual $N_k^{NH}$ do not differ significantly, we have 
approximately 
\be
S_n \approx \frac{1}{\sqrt{n}} \sum_{i = 1}^n S_i. 
\nonumber
\ee
The asymmetry has opposite signs in domains separated by the lines of
zero asymmetry, and the ultimate sensitivity to the neutrino mass
hierarchy can be estimated as
\be
S=\sum_{i = 1}^n  |S_i| \sqrt{\frac{N_i^{NH}}{ \sum_{k = 1}^n 
N_k^{NH}}}\,. 
\nonumber
\ee
One can first estimate independently the sensitivities of individual
domains with the same sign of asymmetry and then sum over the domains.
The real sensitivity will actually be lower because of (i) the
integration (smearing) over bins with different numbers of events and
significances (which will dilute the significance of the most
significant bins) (ii) the integration (smearing) over parts of
domains which have opposite sign of the asymmetry, (iii) uncertainties
of the other oscillation parameters, (iv) degeneracy of parameters,
(v) systematic errors, {\it etc.}.  We address some of these issues
below.

\subsection{$\nu_\mu-$ like events and the mass hierarchy}

The $\nu_\mu-$events produced by the charged current $\nu_\mu$
interaction are observed as muon tracks accompanied by hadronic
cascades.  For these events the energy of the muon $E_\mu$ and the
direction of its trajectory characterized by the angles $\theta_\mu$
and $\phi_\mu$ as well as the total energy of the hadron cascade $E_h$
can be measured. Using this information one can reconstruct the
neutrino energy:
\be 
E_\nu^{r} \approx  E_\mu + E_h - m_N\,,
\nonumber
\ee
where $m_N$ is the nucleon mass.  {In fact the calorimetric (time
integrated) measurement in IceCube provides directly measurement of
$E_\nu$.}

The reconstruction of the neutrino direction is more complicated. In
the first approximation (at sufficiently high energies) one can simply
use $\theta_\nu^r \approx \theta_\mu$ with certain spread which
depends on the neutrino energy. More precise determination is in
principle possible if one uses also the information about the hadron
cascades.  Knowledge of the energy of the cascade narrows down the
uncertainty in the neutrino direction.  Further improvement would be
possible if one determines the plane in which the muon and the
original quark were propagating.  In this plane one can introduce the
angle between the muon and neutrino trajectories, $\theta_\nu$, as
well as the angle between the directions of the quark and neutrino
momenta $\theta_q$.  Then, using the energy and momentum conservation
laws and excluding $\theta_q$, one obtains the expression for the
reconstructed neutrino angle:
\be
\cos \theta_\nu^r \approx \frac{E_\nu^{r 2}+E_\mu^2 - E_h^2}{2 E^r_\nu 
E_\mu}\,,
\nonumber
\ee
where we assume that the muons are ultra-relativistic. In turn, the
knowledge of $\theta_\nu$ and of the muon angles $\theta_\mu$ and
$\phi_\mu$ would allow one to reconstruct the neutrino zenith angle
$\theta_z^{r}$: $\theta_z^{r} = \theta_z^{r}
(\theta_\nu, \theta_\mu, \phi_\mu)$.

There is a number of uncertainties in this reconstruction procedure: 
(i) Errors in the measurements of $E_\mu$, $\theta_\mu$ and $\phi_\mu$; 
(ii) the uncertainty in the point of the neutrino interaction 
(i.e. of the beginning of the muon trajectory); 
(iii) the uncertainty in the position of the center of the hadronic shower and  
(iv) the error in the determination of the energy of the hadron shower. 
 
We will describe the uncertainties of reconstruction of the neutrino
parameters by distribution functions for the reconstructed neutrino
energies and zenith angles:
\be
G_E (E_{\nu}^r , E_\nu),  ~~~G_\theta (\theta_{z}^r , \theta_z), 
\nonumber
\ee
where $E_\nu$ and $\theta_z$ are the true energy and zenith angle of 
the neutrinos. The distributions are normalized in such a way that 
\be
\int d y ~G_y (y^r , y)  = 1\,,  ~~~~~ y = E_\nu, \theta_z\,,
\nonumber
\ee
where the integrations are performed within the appropriate ranges of
the parameters.  For $G_y$ we adopt the Gaussian form
\be 
G_y (y , \sigma_y) = \frac{N_y}{\sqrt{2\pi}\sigma_{y}} 
e^{-\frac{y^2}{2 \sigma_{y}^2}}
\nonumber
\ee
where $N_y$ is the normalization constant, and $\sigma_E$ and
$\sigma_\theta$ are the widths of the energy and angular
reconstruction functions, respectively.  Both widths depend on the
neutrino energy. In this way we obtain the unbinned distribution of
events in the ($E^r_\nu - \cos \theta_{z}^r$) plane:
\be
D_{\alpha}(E^r, \cos \theta^r)  =  
\int  d\cos\theta_z  \int dE_\nu~
G_E (E_{\nu}^r , E_\nu)~G_\theta (\theta_{z}^r , \theta_z) 
 ~V_{\rm eff}~ N_\alpha (E_\nu, \cos\theta_z),   
\label{eq:nres1}
\ee
$\alpha = e, \mu$. Then the binned distributions of events are 
\be
N_{ij, \alpha}^{\rm NH}   =  {2\pi N_A T \rho} 
\int_{\Delta_i (\cos\theta_z^r)} d\cos\theta_z^r \int_{\Delta_j (E_\nu^r)} dE_\nu^r
~D_\alpha (E^r, \cos \theta_z^r). 
\label{eq:nres2}
\ee
We will explore the dependence of our results on the widths $\sigma_E$
and $\sigma_\theta$. For the ideal resolution, $G_y (y^r, y)
= \delta(y^r - y)$, we get from (\ref{eq:nres1}) and (\ref{eq:nres2})
the same results as before.

Consider the opposite limit of large widths, $2\sigma_y \gg \Delta y$.
In this case it is worthwhile to interchange the integrations
$dE_\nu^r d\cos\theta_z^r$ and $dE_\nu d\cos\theta_z$. Then for the
box-like distribution functions, $G_y = 1/2\sigma_y$ in the intervals
$y = y^r \pm \sigma_y$, formulas (\ref{eq:nres1}) and (\ref{eq:nres2})
reproduce the results for large bins $\Delta y
\sim 2\sigma_y$. 

For a contained $\nu_\mu$ event (both the vertex and $\mu$ track are
contained within the detector) the error in the reconstructed neutrino
energy scales linearly with energy, i.e.\ $\sigma_{E_\nu} \sim xE_\nu$
below $\sim 100$ GeV \cite{Albuquerque:2001iv}.  The error in
reconstructing the neutrino arrival direction at low energies is
limited from below by the root mean square value of the scattering
angle, $\theta_{RMS} \sim \sqrt{m_p/E_\nu}$, which corresponds to
$17.5^\circ$ at 10 GeV.

\begin{figure}
\includegraphics[trim=0.in 0.in 0.in 0.in, clip, width=4.in]{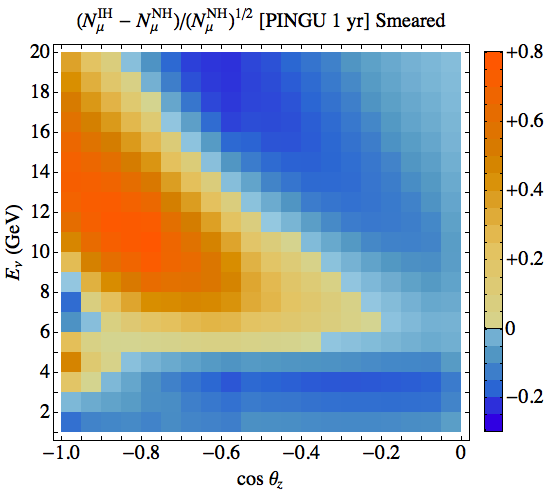}
\caption{
Statistical significance of the determination of the mass hierarchy
after smearing the $\nu_\mu$ events in the ($E^r_\nu$--$\cos\theta^r$)
plane with $\sigma_E = 0.2E_\nu$ and $\sigma_\theta
= \sqrt{m_p/E_\nu}$.}
\label{fig:varyingEtheta}
\end{figure}

\begin{figure}
\includegraphics[trim=0.in 0.in 0.in 0.in, clip, width=4.in]{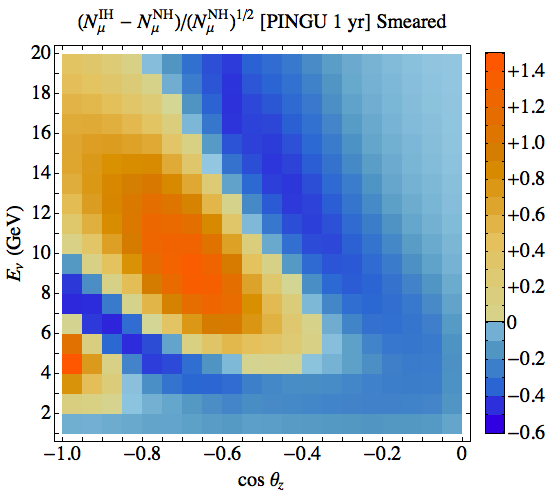}
\caption{
Same as Fig.~\ref{fig:varyingEtheta} but for $\sigma_\theta =
0.5\sqrt{m_p/E_\nu}$.}
\label{fig:varyingEtheta2}
\end{figure}

\begin{figure}
\includegraphics[trim=0.in 0.in 0.in 0.in, clip, width=4.in]{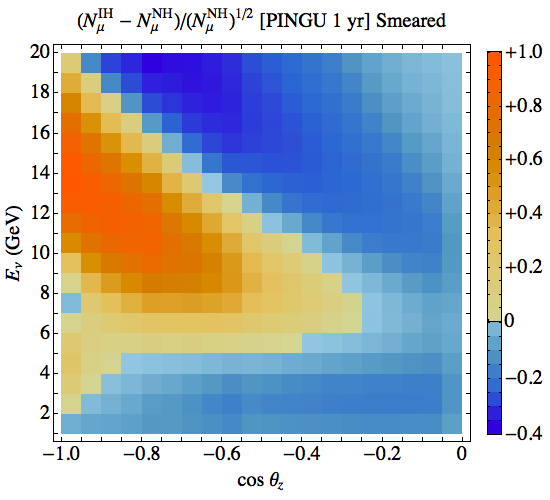}
\caption{Same as Fig.~\ref{fig:varyingEtheta} but 
for $\sigma_E = 2$ GeV.}
\label{fig:varyingEtheta3}
\end{figure}

\begin{figure}
\includegraphics[trim=0.in 0.in 0.in 0.in, clip, width=4.in]{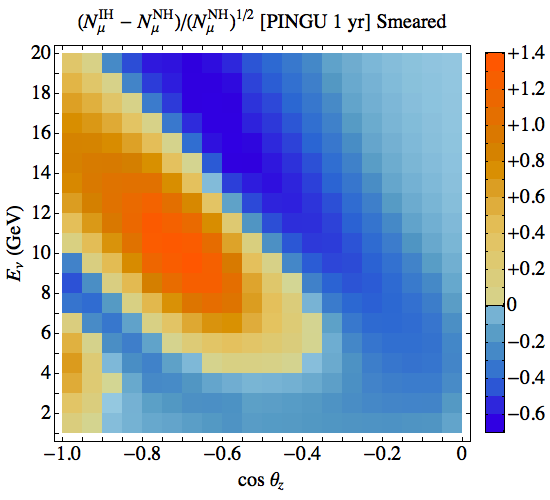}
\caption{Same as Fig.~\ref{fig:varyingEtheta} but with  
$\sigma_E = 2$ GeV and $\sigma_\theta = 0.5\sqrt{m_p/E_\nu}$.}
\label{fig:varyingEtheta4}
\end{figure}

In Figs.~\ref{fig:varyingEtheta} - \ref{fig:varyingEtheta4} we show the 
hierarchy asymmetry in the distributions of the $\nu_\mu$ events 
smeared with energy-dependent Gaussian reconstruction functions 
characterized by different $\sigma_E$ and $\sigma_\theta$.  After 
smearing we integrated the event density over the reconstructed energy 
and zenith angle bins of the size $\Delta (E^r_\nu) = 1$ GeV and 
$\Delta (\cos\theta_z^r) =0.05$.  The smearing leads to a substantial 
decrease of the sensitivity to the neutrino mass hierarchy. This reduction 
is a consequence of the integration over regions with different 
significance and statistics as well as over the regions with different 
signs of the asymmetry.

Considering the effect in each bin as an independent measurement, we 
can find the combined significance as 
\be 
S^{tot} = \sqrt{\sum_{ij} 
S_{ij}^2} = \sqrt{\sum_{ij} \frac{(N_{ij}^{IH} - 
N_{ij}^{NH})^2}{\sigma^2_{ij}}}, 
\nonumber 
\ee 
where the sum over bins can be substituted by the integral. For
illustration we assume that the uncorrelated systematic errors are
proportional to the number of events: $\sigma_{corr} = f N^{NH}_{ij}$,
where $f$ depends on the binning.  We will use $f=5\%$ and $10\%$.  In
general $f$ is a function of neutrino energy and zenith angle.
Therefore the total error in each bin is given by
\be 
\sigma^2_{ij} = N_{ij}^{NH} + (f N_{ij}^{NH})^2. 
\nonumber 
\ee 
Notice that since here the contribution from the systematic error is
proportional to $N^2$, for the same $f$ the role of systematic error
decreases with decreasing size of the bin.

Correlated systematic errors, e.g., those of the overall flux
normalization and of the tilt of the spectrum, apparently can not
reproduce the profile of the distribution of events similar to the
difference of distributions for the two hierarchies.  Therefore, their
effect to a large extent can be reduced to a reduction of the exposure
time and statistics.  Moreover, these correlated systematic errors can
be parameterized and reduced with better measurements of the flux.

\begin{figure}
\includegraphics[trim=0.in 0.in 0.in 0.in, clip, width=5.5in]{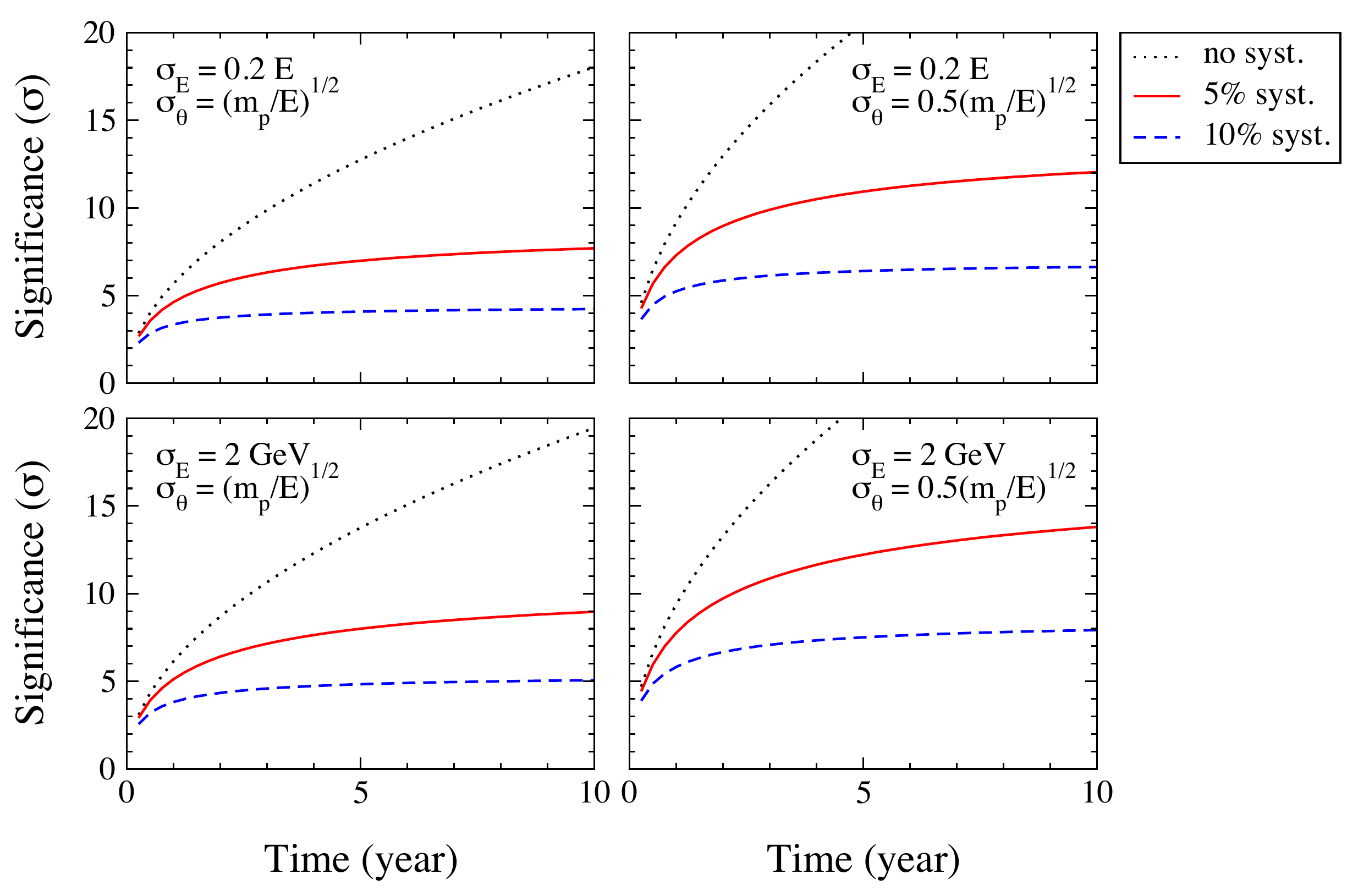}
\caption{
Significance of hierarchy determination vs.\ exposure time for various
smearing schemes illustrated in Figs.~\ref{fig:varyingEtheta},
\ref{fig:varyingEtheta2}, \ref{fig:varyingEtheta3} and 
\ref{fig:varyingEtheta4}.  Different curves correspond to different 
systematic uncertainties ($f$) assumed in addition to statistical 
uncertainties.}
\label{fig:significance}
\end{figure}

Instead of $\theta_\nu$, one could consider the angle $\theta_\mu$
which is measured directly. Smearing over the angle $\theta_\nu$ with
$\sigma_\theta \sim \sqrt{m_p/E_\nu}$ essentially corresponds to the
transition from $\theta_\nu$ to $\theta_\mu$.  Again, due to
measurements of the cascade energy this uncertainty will be further
reduced.  In Fig.~\ref{fig:varyingEtheta} we used $\sigma_E =
0.2E_\nu$ and $\sigma_\theta = \sqrt{m_p/E_\nu}$.  It can be seen from
the figure that the region of the highest significance is between the
lines $E_\nu /{\rm GeV} \approx 20 (|\cos \theta_z | - 0.2)$ and
$E_\nu /{\rm GeV} \approx 15 (|\cos \theta_z | - 0.2)$, and between
$\cos\theta_z = -1$ to $-0.7$.  It is shifted towards higher energies
compared to the un-smeared case in Fig.~\ref{fig:asym-mu}.

Detection of hadronic cascades can in principle improve the
determination of the neutrino angle. In Fig.~\ref{fig:varyingEtheta2}
we therefore use the angular resolution $\sigma_\theta =
0.5\sqrt{m_p/E_\nu}$.  Notice that with better angular resolution the
region of high sensitivity shifts to lower energy and shallower zenith
angle bins, approximately along the lines of high significnace.

To study the effect of the energy dependence of $\sigma_E$ on the
neutrino mass hierarchy determination, in
Fig.~\ref{fig:varyingEtheta3} we use the fixed energy resolution
$\sigma_E = 2$ GeV and the angular resolution $\sigma_\theta =
\sqrt{m_p/E_\nu}$. A comparison with Fig.~\ref{fig:varyingEtheta} 
shows that fixed $\sigma_E$ leads to a somewhat higher significance of
the hierarchy determination.

In Fig.~\ref{fig:varyingEtheta4} we used the energy resolution
$\sigma_E = 2$ GeV but the angular resolution $\sigma_\theta =
0.5\sqrt{m_p/E_\nu}$. Figs.~\ref{fig:varyingEtheta4}
and~\ref{fig:varyingEtheta2} show that with improving angular
resolution the significance of the neutrino mass hierarchy
determination increases significantly.

Figure \ref{fig:significance} shows the significances of the hierarchy
identification, $S^{tot}$, for different smearing schemes described
above and for different uncorrelated systematics ($f$). The 5-year
significances from the plots can be compared with the results found
when no smearing is performed: $S^{tot} = 45.5\sigma$ (no
systematics), $S^{tot} = 28.9\sigma$ ($f = 5\%$) and $S^{tot} =
18.8\sigma$ ($f = 10\%$).  Note that if IH is the true hierarchy, then
in the first approximation the results simply correspond to
Figs.~\ref{fig:varyingEtheta} -- \ref{fig:varyingEtheta4} with
reversed signs of the asymmetry, although significances become
somewhat lower in general.

\subsection{Effects of parameter degeneracy}

In our computations of the hierarchy asymmetry we used the fixed
values of $\Delta m_{32}^2$, $\theta_{13}$, $\theta_{23}$ and
$\delta$.  In certain kinematic regions, uncertainties in these
parameters may lead to the same effects as the hierarchy change.  That
is, the difference of the event distributions for the true and assumed
values of the parameters may have the same pattern as the difference
of event distributions for the normal and inverted mass
hierarchies. The simplest way to reduce the degeneracy is to select
only parts of the $E_\nu - \cos\theta_z$ space where the effect of
hierarchy change dominates over other effects.

1. As follows from the comparison of
Figs.~\ref{fig:event-cp05}, \ref{fig:event-cp025}, \ref{fig:event-cp052}
and Fig.~\ref{fig:asym-mu}, the effect of CP-phase $\delta$ at high
energies is smaller than $10\%$ of the hierarchy effect, and so it can
be neglected in the first approximation. Also, it is characterized by
a different pattern, and therefore one can select the regions in the
$E_\nu - \cos\theta_z$ plane in such a way as to further suppress the
effect of the CP-phase (e.g. the selected bins should include domains
with different signs of CP-asymmetry).  We computed the hierarchy
asymmetry for $\delta = \pi/2$ and found that indeed the CP effect can
be neglected in the first approximation.

\bfig
\includegraphics[trim=0.in 0.in 0.in 0.in, clip,width=4.in]{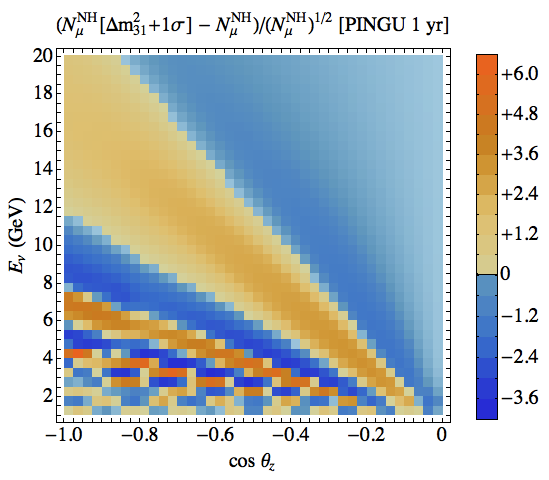}
\caption{
Effect of uncertainty in $\Delta m_{31}^2$.  Shown is the difference
of numbers of events for $\Delta m_{31}^2$ and $\Delta m_{31}^2 +
1\sigma = 2.47\cdot 10^{-3}$~eV$^2$.  NH is assumed.  }
\label{fig:dms}
\efig

2.  Effects of the uncertainty in $\theta_{23}$ are strong in the
region along the line $E_\nu/{\rm GeV} \approx 23 |\cos \theta_z|$
(Fig.~\ref{fig:event23mu}). In the high energy part of this region, $E
> 10$ GeV, the hierarchy asymmetry is small. There is still an overlap
of the regions of large effects of the hierarchy and $\theta_{23}$
uncertainty at low energies. The bins can be selected in such a way
that the effect of the uncertainty in $\theta_{23}$ substantially
cancels out.  Furthermore, one expects some improvements in the
determination of $\theta_{23}$ from the accelerator and combination of
reactor and accelerator experiments, so that the degeneracy will be
further reduced.  In particular, combining the reactor data with the
forthcoming results from T2K and NOvA will allow us to determine
$\sin^2 \theta_{23}$ with an accuracy between about 0.04 (for maximal
2-3 mixing) and 0.008 (for
$\sin^2 \theta_{23}=0.4$) \cite{Huber}. This has to be compared with
the current uncertainty of this parameter
$\delta(\sin^2 \theta_{23})\simeq 0.05$ \cite{Fogli} (all numbers
correspond to $1\sigma$).

3.  The effect of uncertainty in $\Delta m_{31}^2$ is illustrated in
Fig.~\ref{fig:dms}, where we show the difference of numbers of events
for the best fit and shifted upwards by $1 \sigma$ values of $\Delta
m_{31}^2$. (A downward shift by $1 \sigma$ switches signs in
Fig.~\ref{fig:dms}, although with somewhat smaller significances.)  In
the limit $\Delta m_{21}^2 = 0$ variations of $\Delta m_{31}^2$ are
equivalent to a corresponding shift of the oscillatory pattern in the
energy scale (see Fig.~\ref{fig:probab2}) at high energies ($E > 8$
GeV). Note that the shift is different for different zenith angles and
for neutrinos and antineutrinos. The regions of the strongest effect
of this shift on the hierarchy asymmetry have substantial overlap with
the regions of strong hierarchy asymmetry, and therefore certain
selection of the integration regions (binning) is required in order to
reduce the degeneracy and disentangle the two effects. Clearly,
further improvements of the accuracy of measurements of $\Delta
m_{31}^2$ by MINOS, T2K and NOvA will alleviate this problem.  In
particular, T2K and NOvA will measure $\Delta m_{31}^2$ with an
accuracy of $5\times 10^{-5}$ eV$^2$ ($1\sigma$) \cite{Huber}.  This
is about a factor of two better than the current uncertainty of this
parameter \cite{Fogli}.

To further explore if the uncertainty in $\Delta m_{31}^2$ can mimic
the ``wrong'' hierarchy, we applied the same $\sigma_E = 0.2E_\nu$ and
$\sigma_\theta = \sqrt{m_p/E_\nu}$ smearing as in
Fig.~\ref{fig:varyingEtheta} to the best-fit and $+1\sigma$ deviation
of the $\Delta m_{31}^2$ values.  The resulting significances are
plotted in Fig.~\ref{fig:dmssmeared}.  Note that the region of the
highest sensitivity to $\Delta m_{31}^2$ has shifted to higher
energies and is concentrated in a narrower, $\cos\theta_z < -0.8$,
range as compared to the region of the highest significance of the
hierarchy asymmetry (cf. Figs.~\ref{fig:dmssmeared}
and~\ref{fig:varyingEtheta}).

\bfig 
\includegraphics[trim=0.in 0.in 0.in 0.in, clip, width=4.in]{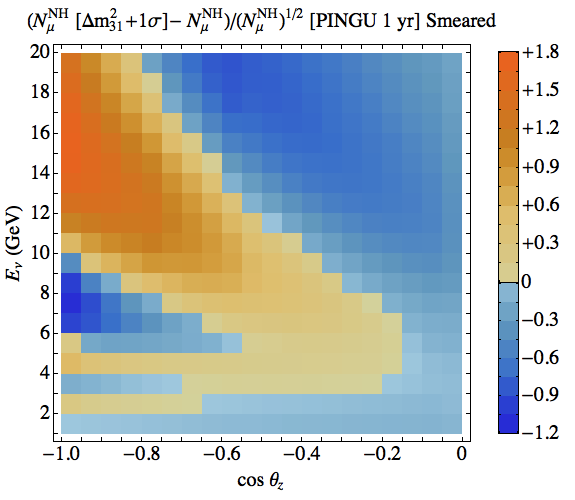} 
\caption{Same as in Fig.~\ref{fig:dms} but after smearing the $\nu_\mu$ 
events in the ($E^r_\nu$--$\cos\theta^r$) plane with $\sigma_E = 0.2E_\nu$ 
and $\sigma_\theta = \sqrt{m_p/E_\nu}$. }
\label{fig:dmssmeared} 
\efig 

\bfig
\includegraphics[trim=0.in 0.in 0.in 0.in, clip, width=4.in]{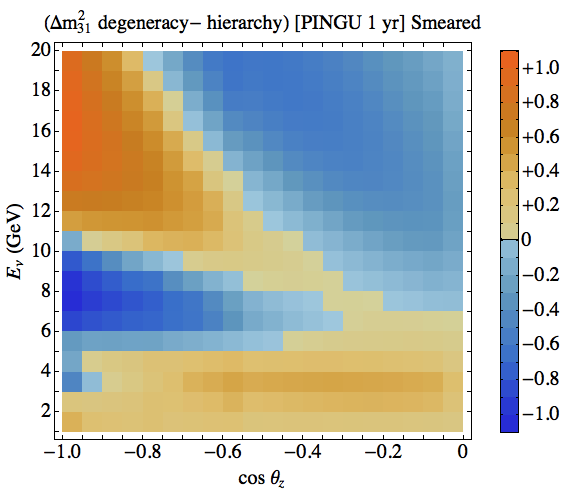}
\caption{
Difference of significances in the ($E^r_\nu$--$\cos\theta^r$) plane
between the $\Delta m_{31}^2$ uncertainty (Fig.~\ref{fig:dmssmeared})
and the mass hierarchy (Fig.~\ref{fig:varyingEtheta}).}
\label{fig:Dm31hiearchy}
\efig

Figure~\ref{fig:Dm31hiearchy} shows the difference between the
significances plotted in Figs.~\ref{fig:dmssmeared}
and \ref{fig:varyingEtheta}.  Notice that the significances of the
hierarchy determination in the region $E_\nu\simeq 6-13$ GeV are
reduced considerably. The effect of the degeneracy can be suppressed
by making summation of significances only over certain domains in the
$E_\nu - \cos\theta_z$ plane.

To estimate the effect of the uncertainty of the value of $\Delta
m^2_{31}$ on the significance of the hierarchy determination, we
simulated the data for NH $N_\mu^{NH}$ for a fixed ``true'' value of
$\Delta m^2_{31, true}$ and then fitted them with IH, treating $\Delta
m^2_{23}$ as a free parameter.  We then minimized $S^{tot}$ with
respect to $\Delta m_{32}^2$ and found the corresponding values of
$\Delta m^2_{23, fit}$ and $S^{tot}_{min}$. Next, we repeated the same
procedure for different true values of $\Delta m^2_{31, true}$ within
its $1\sigma$ allowed range.  This procedure is illustrated in
Fig.~\ref{fig:Dm31minimized}. The left panel shows $S^{tot}$ versus
$\Delta m^2_{32,fit}$ for $\Delta m^2_{31,true} = 2.35\cdot 10^{-3}$
eV$^2$ (vertical line). In the right panel of
Fig.~\ref{fig:Dm31minimized} we show the values of $S^{tot}_{min}$
obtained through this procedure as functions of $\Delta m^2_{31,
true}$ (solid lines). For comparison we show also $S^{tot}$ for
$\Delta m^2_{23, fit}=\Delta m^2_{31, true}$, i.e.\ without variations
of $\Delta m^2_{23, fit}$ (dashed lines).  As follows from the figure,
variation of $\Delta m^2_{23, fit}$ reduces the significance of the
hierarchy identification $S^{tot}$ by $\sim 50 \%$, and this reduction
weakly depends on $\Delta m^2_{31, true}$.  The values $\Delta m_{23,
fit, min}^{2}$ are within the present $2\sigma$ uncertainties of
determination of this mass difference ($2.17\cdot 10^{-3} - 2.59\cdot
10^{-3}$) eV$^2$ \cite{Fogli:2012ua}. To calculate the significances
presented in the figure we have smeared the event distributions with
$\sigma_E = 0.2E_\nu$ and $\sigma_\theta = \sqrt{m_p/E_\nu}$ as in
Fig.~\ref{fig:varyingEtheta}.

\bfig
\includegraphics[trim=0.in 0.in 0.in 0.in, clip, width=5.5in]{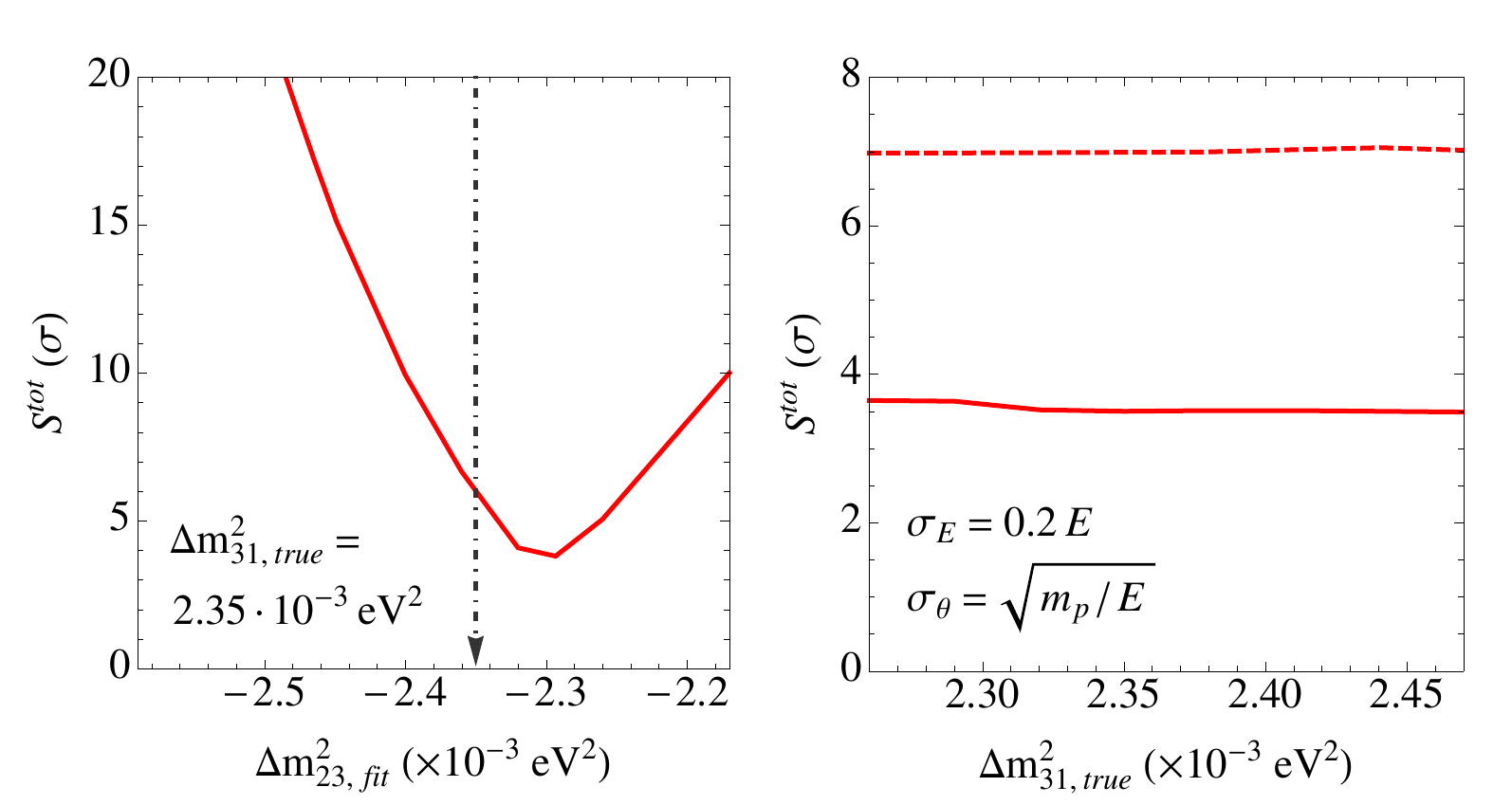}
\caption{
{\em Left panel--} Significance of determination of the hierarchy as a
function of the ``fit'' values of $\Delta m_{23, fit}^2$ when the
``true'' value of $\Delta m_{31, true}^2 = 2.35\cdot 10^{-3}$ eV$^2$
(vertical line). {\em Right panel --} Significances as functions of
$\Delta m_{31,true}^2$ after minimization as in the left panel (solid
lines) and when $\Delta m_{23, fit}^2=\Delta m_{31,true}^2$ (dashed
lines).  A systematic uncertainty $f = 5\%$ and 5-yr PINGU data was
used.}
\label{fig:Dm31minimized}
\efig

\section{Conclusions} 

The main goal of our paper was to attract attention to the possibility
of determination of the neutrino mass hierarchy with huge atmospheric
neutrino detectors like PINGU, to propose a method of quick estimation
of the sensitivity of such detectors to the mass hierarchy and to
outline challenges on the way of realization of this idea.

1. After the determination of the leptonic 1-3 mixing angle, the
structure of the neutrino oscillograms discussed in refs.~\cite{our2}
and \cite{our3} (see also Fig.~\ref{fig:oscillogr}) is well
determined, and the position of the main structures in the $E_\nu
- \cos \theta_z$ (kinematic) plane is fixed.  For $E_\nu > 1$ GeV
these include the MSW resonance in the Earth's mantle domain as well
as the MSW resonance and the three parametric ridges in the core
domain of the oscillograms.

2. The multi-megaton ice (water) detectors like PINGU will allow one
to reconstruct the oscillograms and determine yet unknown neutrino
parameters: the mass hierarchy (the sign of $\Delta m^2_{31}$), the
deviation of the 2-3 mixing from the maximal one, and in principle,
the CP-violation phase as the next step. In addition, once the
neutrino mass hierarchy has been established, the data from
multi-megaton detectors should allow a significantly better
determination of the value of $\Delta m_{31}^2$.

3. At the probability level the effects of the change of the mass
hierarchy and of variations of $\theta_{23}$ can be of order
1. However, there are several factors that substantially reduce the
effects at the level of observable events. We have identified and
studied in detail the following factors:

a) $\nu - \bar{\nu}$ summation, which is related to the presence of
both neutrinos and antineutrinos in the original neutrino flux. The
hierarchy asymmetry survives due to a factor of $\sim$2 difference of
the neutrino and antineutrino cross-sections as well as some
difference of the original neutrino fluxes.

b) Flavor screening, which leads to the suppression factors like
$(s_{23}^2 r - 1)$ and $(s_{23}^2 \bar{r} - 1)$ and is related to the
presence of both $\nu_e$ and $\nu_\mu$ in the original flux.

c) Dilution of the significance: often large significances of the
hierarchy asymmetry and other event number differences appear in bins
where numbers of events are small. Then summation of signals in these
high-significance bins and in bins where the numbers of events are
larger but the significance is lower leads to a dilution of the high
significance.

d) Smearing: finite energy and angular resolutions mean that
integrations over rather large domains in the $E_\nu - \theta_z$ plane
have to be performed. These domains usually contain regions with
different signs of the considered effect (e.g., hierarchy
asymmetry). Therefore the integration leads to a significant
suppression of the studied effects.

e) Parameter degeneracies.

4. We presented the significance plots for the determination of the
neutrino mass hierarchy, deviation of the 2-3 mixing from the maximal
one and CP-phase.

5. To evaluate the possibility to establish the neutrino mass
hierarchy, we performed smearing of the event number distributions
using Gaussian functions for reconstructing the true neutrino energies
and zenith angles.  We have studied the dependence of the
significances integrated over certain ranges of $E_\nu$ and
$\cos\theta_z$ on the widths of the reconstruction functions. Our
preliminary estimates show that after 5 years of PINGU 20 operation
the significance of the determination of the hierarchy can range from
$\sim 3 \sigma$ to $10 \sigma$ (with parameter degeneracies taken into
account), depending on the accuracy of reconstruction of the neutrino
energy and zenith angle.

6. The smearing procedure implemented in this paper captures the main
uncertainties of reconstruction of the true neutrino energies and
zenith angles rather accurately. The smearing we have adopted gives a
good estimate (at least at this stage of knowledge of future
experimental characteristics) of the accuracy of reconstruction of the
neutrino parameters. By varying the smearing parameters in rather wide
ranges we covered essentially all possibilities of practical interest
and explored the detecor resolutions necessary to achieve a given
significance of the hierarchy determination.

7. The parameter degeneracy effects can be significant, so that
similar patterns of event distribution in the $E_\nu - \theta_z$ plane
or in its parts can be obtained due to either changing the neutrino
mass hierarchy or due to variations of the 2-3 mixing or of $\Delta
m_{31}^2$ within their currently allowed ranges. For each parameter we
identified the kinematic regions of the smallest degeneracy (where the
effect we are interested in dominates). The effects of the parameter
degeneracy can be reduced by selecting particular regions of
integration over $E_\nu$ and $\cos\theta_z$. In addition, forthcoming
measurements of neutrino parameters (in MINOS, T2K, NOvA and in
reactor experiments) should provide us with more accurate values of
these parameters and further reduce the effects of parameter
degeneracy.

\section*{Acknowledgments}

We are grateful to D.~F.~Cowen, D.~Grant, J.~Koskinen and E.~Resconi
for correspondence on the design and performance of the PINGU
detector.  We also thank M.~Blennow, S.~Choubey, J.~G.~Learned and
T.~Schwetz for useful comments and discussions, and M.~Ribordy for
pointing out an error in the smearing process.  E.A. and A.S. thank
the Galileo Galilei Institute for Theoretical Physics for the
hospitality and the INFN for partial support during the work on a
revised version of this paper. S.R. thanks the Abdus Salam
International Centre for Theoretical Physics, where this work was
initiated, for hospitality.  A.S. acknowledges support by the
Alexander von Humboldt Foundation and is grateful to the MPI f\"ur
Kernphysik, Heidelberg, where a part of this work has been done, for
hospitality.

\section*{Appendix. Proof of the relation $\bar{P}_{\alpha\beta}=
P_{\alpha\beta}(\Delta \to -\Delta)$ in the limit $\Delta m_{21}^2\to 0$}

\noindent
Let us prove that in the limit $\Delta m_{21}^2\to 0$ the oscillation
probabilities for antineutrinos are given by those for neutrinos with
the substitution $\Delta m_{31}^2 \to -\Delta m_{31}^2$.  A slightly
different proof of this statement can be found in
\cite{minakata}. 

The evolution equation for the neutrino state vector in the propagation 
basis is 
\be
i\frac{d}{dx}\tilde{\nu}=\tilde{H}(x)\tilde{\nu}\,,
\label{eq:evol}
\ee
where $\tilde{\nu}=(\nu_e, \nu_{\tilde{2}}, \nu_{\tilde{3}})$ and the 
effective Hamiltonian $\tilde{H}(x)$ in the limit $\Delta m_{21}^2\to 0$  
takes the form 
\be
\tilde{H}(x)=\left(
\begin{array}{ccc}
s_{13}^2 \Delta+V(x) &0 & s_{13}c_{13}\Delta\\
0 & 0 & 0\\
s_{13}c_{13}\Delta &0 & c_{13}^2 \Delta 
\end{array}
\right).
\label{eq:H1}
\ee
Here $\Delta\equiv \Delta m_{31}^2/2E$. 
The evolution matrix (the matrix of the transition amplitudes) then has 
the form 
\be
A(x)=\left(
\begin{array}{ccc}
A_{ee} &0 & A_{e\tilde{3}}\\
0 & 1 & 0\\
A_{\tilde{3}e} &0 & A_{\tilde{3}\tilde{3}} 
\end{array}
\right)
\label{eq:A1}
\ee
From unitarity of this matrix it follows that 
$|A_{\tilde{3}e}|=|A_{e\tilde{3}}|$. 
For antineutrinos, one has to flip the sign of the potential $V(x)$ in 
eq.~(\ref{eq:H1}). Obviously, this is equivalent to flipping the sign 
of $\Delta$ in (\ref{eq:H1}) and additionally changing the overall 
sign of the Hamiltonian $\tilde{H}(x)$. The latter can be compensated 
by complex conjugating the evolution equation (\ref{eq:evol}). Thus, we find 
\be
A(\bar{\nu}, \Delta)=A^*(\nu, -\Delta)\,.
\label{eq:relation1}
\ee
Recall now that in the limit $\Delta m_{21}^2\to 0$ the probabilities 
of various flavour transitions are expressed only through 
$P_A=|A_{e\tilde{3}}|^2$ and $\cos\phi_X$ where $\phi_X=
arg(A_{\tilde{2}\tilde{2}} A_{\tilde{3}\tilde{3}}^*) =arg(A_{\tilde{3}
\tilde{3}}^*)$ (see eqs.~(\ref{eq:probee})-(\ref{eq:probmutau})). 
Therefore we find that the oscillation probabilities for antineutrinos are 
given by those for neutrinos of the opposite neutrino mass hierarchy 
({\it i.e.} \ with $\Delta \to -\Delta$).

\end{document}